%% file: main.tex
\pdfoutput=1 
\documentclass[11pt,a4paper]{article}

\usepackage{jheppub} 

\usepackage[T1]{fontenc} 
 
\usepackage{amsmath,amsthm,amssymb}
\usepackage{multicol}
\usepackage{soul}
\usepackage{blkarray}
\usepackage{array}

\newcommand{\Z}{\mathbb{Z}}

\newcommand{\imrho}{\text{Im}\rho}

\newcommand{\V}{\mathcal{V}}
\newcommand{\antiD}{\overline{\text{D3}}}
\newcommand{\Op}[1]{\mathcal{O}\left(#1\right)}
\newcommand{\bigo}[1]{\mathcal{O}(#1)}

\newcommand{\blue}[1]{\textcolor{blue}{#1}}

\newcommand{\zGKP}{\zeta_{\rm GKP}}
\newcommand{\cD}{c_{\rm D3}}

\numberwithin{equation}{section}

\usepackage[utf8]{inputenc}
\usepackage[english]{babel}
\usepackage{textcomp}
\usepackage{amsmath}
\usepackage{mathtools}
\usepackage{gensymb}
\usepackage{multirow}
\usepackage{multicol}
\usepackage{amsbsy}
\usepackage{amssymb}
\usepackage{capt-of}
\usepackage[T1]{fontenc} 
\usepackage{physics}
\usepackage{multicol} 
\usepackage{multirow}
\usepackage[table]{xcolor}
\usepackage{lscape}
\usepackage[format=plain,small,labelfont={bf},up]{caption} 
\usepackage[title]{appendix}

\def\be{\begin{equation}}
\def\ee{\end{equation}}
\def\bea{\begin{eqnarray}}
\def\eea{\end{eqnarray}}

\def\ni{\noindent}

\usepackage{booktabs} 
\usepackage{float} 

\usepackage{hyperref} 
\usepackage{graphicx} 
\usepackage{epstopdf} 
\usepackage{paralist} 
\usepackage{subcaption}
\usepackage[framemethod=TikZ]{mdframed}

\RequirePackage{color}
\usepackage{colortbl}
\definecolor{denim}{rgb}{0.08, 0.38, 0.74}

\hypersetup{
    colorlinks=true,
    linkbordercolor={white},
    linkcolor={denim},
    citecolor={denim},
    filecolor={denim},      
    urlcolor={denim},
}

\title{\huge De Sitter vacua -- when are `subleading corrections' really subleading?}

\author[a]{Bruno Valeixo Bento,}
\author[b]{Dibya Chakraborty,}
\author[a]{Susha Parameswaran,}
\author[c]{Ivonne Zavala}

\affiliation[a]{Department of Mathematical Sciences, University of Liverpool, Liverpool L69 7ZL, UK}
\affiliation[b]{Department of Physics, Ashoka University,
Plot 2, Rajiv Gandhi Education City, P.O.~Rai, Sonipat 131029, Haryana, India
}
\affiliation[c]{Department of Physics, Swansea University, Singleton Park,  SA2 8PP, UK}

\emailAdd{Bruno.Bento@liv.ac.uk}
\emailAdd{dibya.chakraborty@ashoka.edu.in}
\emailAdd{susha@liv.ac.uk}
\emailAdd{e.i.zavalacarrasco@swansea.ac.uk}

\abstract{We consider various string-loop, warping and curvature corrections that are expected to appear in type IIB moduli stabilisation scenarios.  It has recently been argued, in the context of strongly-warped LVS de Sitter vacua, that it is impossible to achieve parametric suppression in all of these corrections simultaneously \cite{Junghans:2022exo}.  We investigate corrections in the context of the recently discovered weakly-warped LVS de Sitter vacua, which represent a distinct branch of solutions in type IIB flux compactifications. The weakly-warped solution is supported by small conifold flux numbers $MK \lesssim 32$, but still requires a large flux contribution to the D3-tadpole, now from the bulk. Warping corrections become less problematic, and some corrections even help to reach the weakly-warped regime of parameter space.  Other corrections continue to be dangerous and would require numerical coefficients to be computed --- and found to be small --- in order not to destroy the consistency of the weakly-warped LVS de Sitter solution. We motivate why this may be possible. 
}

\begin{document}
\maketitle

\input{Sections/introduction}

\input{Sections/potential}

\input{Sections/DeformationModulusStabilisation}
\input{Sections/LVS}
\input{Sections/corrections}
\input{Sections/conclusions}

\section*{Acknowledgements}
We are grateful to Ignatios Antoniadis, Michele Cicoli, Arthur Hebecker, Daniel Junghans, Severin L\"ust, Erik Plauschinn, Simon Schreyer and Gerben Venken for discussions. 
IZ is partially supported by STFC, grant  ST/T000813/1. 

\smallskip
\noindent For the purpose of open access, the authors have applied a Creative Commons Attribution (CC BY) licence to any Author Accepted Manuscript version arising. Data access statement: no new data were generated for this work.

\bibliographystyle{utphys}

\bibliography{references}

\end{document}

%% file: Sections/introduction.tex
\section{Introduction}
\label{sec:introduction}

The microscopic nature of the Dark Energy that dominates our Universe today and drives its accelerated expansion remains arguably the biggest question in fundamental physics today.  So far, cosmological observations are consistent with Dark Energy being a tiny, positive cosmological constant, or vacuum energy, sourcing a de Sitter Universe.  The string theory landscape could in principle provide an explanation for this tiny, positive cosmological constant, with its prediction of a multitude of four-dimensional Universes, each one with a different total vacuum energy, including ones that match the observed value.  However, this landscape relies on de Sitter vacua obtained with a delicate balance of tree-level and leading-order perturbative and/or non-perturbative corrections in two expansions; using weak string coupling and large volume.  There is an ongoing fruitful debate on the validity of these solutions, with constructions becoming ever-more explicit and tests on their robustness ever-more rigorous.  Any candidate construction must also face the fundamental issues around de Sitter vacua, such as how to formulate an S-matrix in spacetimes with cosmological horizons. 

This paper is inspired by \cite{Junghans:2022exo}, which made a comprehensive analysis of the various subleading corrections to type IIB moduli stabilisation scenarios that have been considered in the literature and assessed how dangerous they are to strongly-warped Large Volume Scenario de Sitter vacua, with light volume and conifold moduli. Ref. \cite{Junghans:2022exo} uncovered a so-called `non-perturbative no-scale' behaviour, whereby several curvature, warping and $g_s$ corrections --- although parametrically suppressed in the off-shell potential --- appear enhanced in the moduli vevs and vacuum energy.  Moreover, it was found that some contributions have no parametric suppression at all; the ten-dimensional constructions must therefore be such that these contributions are vanishing or suppressed for other reasons.  Finally, by taking the specific model presented in \cite{LVSdS:2010.15903} as a case study, and assuming that the latter unsuppressed contributions could be neglected, Ref.\cite{Junghans:2022exo} found that the remaining corrections could not be simultaneously parametrically suppressed in small $g_s$ and large $\mathcal{V}$.  Thus, for a consistent de Sitter vacuum, the numerical coefficients of these corrections would have to be computed, and their suppression would have to be arranged via the topological choices in the ten-dimensional construction.  

These issues have been further studied in \cite{Gao:2022fdi}, which formulated the so-called ``Parametric Tadpole Bound'', by working with a particular combination of corrections and showing that it could be suppressed in the presence of a large D3-tadpole.   Ref. \cite{Junghans:2022kxg}  extended this analysis by including the full set of corrections analysed in \cite{Junghans:2022exo} to derive a similar bound, in terms of the D3-tadpole and the topological data of the Calabi-Yau and branes/orientifolds.  It was again found that large tadpoles could suppress all the corrections, with numbers needed ranging from $\mathcal{O}(500)$ to $\mathcal{O}(10^6)$ or more, depending on the other topological data.  To summarize, in order for strongly-warped LVS dS vacua to survive all corrections, the topological choices in the 10d compactification must be such that some of their numerical coefficients happen to be small and/or large tadpoles can be accommodated. To establish whether the strongly-warped LVS dS vacua are in the string landscape or in the `string theory swampland', one would need constructions with large tadpoles and/or computational control over all the potentially dangerous subleading corrections, which may well be developed in the coming years (see e.g. \cite{Cicoli:2021rub} for a recent study on $\alpha'$ corrections and \cite{Crino:2022zjk} for recent results on Calabi-Yau's allowing for large D3-tadpoles).

At the same time, it is interesting to ask if similar issues arise for other proposed de Sitter scenarios.  In this paper, we revisit the recently discovered \emph{weakly-warped} de Sitter vacua \cite{Bento:2021nbb}.  This construction is related to the Large Volume Scenario, but emerges when considering a previously unexplored regime of parameter space, where the Klebanov-Strassler solution is only very weakly warped and there is only a mild hierarchy between UV and IR scales.   It is thus associated with small flux numbers threading the conifold, $MK \lesssim 32$.  It should be emphasised that the weakly-warped LVS solutions are more than just a different regime of parameter space of the strongly-warped LVS solutions; their form in terms of the moduli --- in particular the conifold deformation modulus --- are distinct, as different terms dominate in the warped K\"ahler potential that mixes the volume and conifold moduli.

Strongly-warped KKLT and LVS constructions are vulnerable to runaway instabilities after uplifting anti-de Sitter vacua to de Sitter \cite{upliftingrunaways2019, Crino:2022sey}.  Avoiding a runaway volume modulus consistently with the supergravity expansion $g_s M \gg 1$ requires large conifold flux numbers, $MK \gtrsim \mathcal{O}(1000)$ and $\mathcal{O}(100)$ for KKLT and LVS respectively, again leading to a tadpole problem \cite{upliftingrunaways2019, Crino:2022sey, tadpoleProblem}.  The same was argued to hold true for the conifold modulus in \cite{upliftingrunaways2019, Crino:2022sey}, although \cite{Lust:2022xoq} subsequently suggested that higher dimensional constraint equations, which are important when warped-down light KK modes are not included in the 4d low-energy effective field theory, may help evade the conifold instability. The weakly-warped LVS scenario instead evades both volume and conifold runaways via large $AW_0$, where $W_0$ is the bulk superpotential arising from the stabilised complex structure moduli and the dilaton that have been integrated out and $A$ is the coefficient of the non-perturbative contribution. Interestingly, the requirement for a large $W_0$ pushes the flux contribution to the D3-tadpole towards larger values, putting the weakly-warped LVS in a position similar to the strongly-warped LVS as far as the need for large tadpoles is concerned \cite{Gao:2022fdi,Junghans:2022kxg}. On the other hand, the weakly-warped LVS solution has no bulk-singularity problem \cite{Carta_2019, Gao:2020xqh}, and no problem with the conifold fitting into the bulk \cite{Carta:2021lqg}.  Also, the warped KK scale is greater than the mass of the conifold and volume moduli, making the solution consistent with the Kaluza-Klein truncation.  

The purpose of this paper is to explore the robustness of this weakly-warped LVS de Sitter solution against subleading curvature, warping and $g_s$ corrections.  We will find that the weak warping renders some the corrections less dangerous, compared to the strongly-warped LVS case.  Some of the corrections even help in bringing us into the weakly-warped regime of parameter space.  Other corrections, however, continue to be dangerous, representing contributions to the off-shell scalar potential, moduli vevs and vacuum energy that have no parametric suppression in small $g_s$ or large volume.  Our results are summarised in Table \ref{tb:summary}.  In order for the weakly-warped dS solution to be consistent, with the neglected corrections genuinely suppressed, the higher dimension construction must be such that certain numerical coefficients are small or even vanishing; as we will discuss in the closing section, this may indeed be possible. 

The paper is organised as follows.  In Sections \ref{sec:ww-conifold}--\ref{sec:ww-LVS-dS} we review the weakly-warped de Sitter vacua, in a presentation that supersedes that in\footnote{We correct an error in the original presentation \cite{Bento:2021nbb}.} \cite{Bento:2021nbb}.  We begin with a description of warped compactifications and the warped deformed conifold, and the 4d low energy effective field theory for the volume and conifold moduli.  Then we discuss the conifold modulus stabilisation and finally we treat the full volume and conifold moduli stabilisation in the context of a weakly-warped LVS scenario.  The main part of the paper is Section \ref{sec:dangerous-corrections}, where we compute the subleading corrections to the off-shell scalar potential, and then examine their consequences for the moduli vevs and vacuum energy.  We conclude with a summary and outlook in Section \ref{sec:conclusions}.

%% file: Sections/potential.tex
\section{The weakly-warped deformed conifold}
\label{sec:ww-conifold}

In this section we briefly summarise the main ingredients of  type IIB flux compactifications with warped throats and their 4d effective field theory description. In particular, we are mostly interested  in the volume modulus and the deformation modulus of a Klebanov-Strassler (KS) warped deformed conifold. 

\subsection{Warped compactifications and the volume modulus}  

We  start from the type IIB supergravity bosonic action in the Einstein-frame given by\footnote{In our conventions, the string  and Einstein frame 10d metrics are related by $G_{MN}^S = e^{\frac{\phi-\phi_0}{2}}G_{MN}^E$, where $e^{\phi_0}=g_s$ with $\phi_0$ being the background value of the dilaton \cite{ValeixoBento:2023afn}. This means that $\kappa = \kappa_{10} g_s$, hence $2\kappa^2 = (2\pi)^7g_s^2\alpha'^4$, and  volumes are frame independent. \label{F:vev_shift}}
\begin{align}
    S_{IIB}^{boson} &= \frac{1}{2\kappa^2}\int d^{10}x \sqrt{-g_{10}} \left\{R_{10} - \frac{\partial_M\tau\partial^M\Bar{\tau}}{2(\text{Im}\tau)^2} - \frac{g_s|G_3|^2}{2(\text{Im}\tau)} - \frac{g_s^2|F_5|^2}{4} \right\} \nonumber \\
    &\quad - \frac{ig_s^2}{8\kappa^2}\int \frac{C_4\wedge G_3\wedge \overline{G}_3}{\text{Im}\tau} \,,
    \label{eq:TypeIIB}
\end{align}
where $\tau = C_0+ie^{-\phi}$ is the axio-dilaton,  $G_3 = F_3 - \tau H_3$ is the 3-form flux, and $F_5=dC_4 - \frac12C_2\wedge H_3+\frac12 B_2\wedge F_3$ with $C_0, C_2, C_4$ the RR potentials.

We consider a flux compactification, whose internal compact space consists of a finite portion of a warped throat described by the warped non-singular deformed conifold \cite{KS2000supergravity},  glued to a compact Calabi-Yau.  
The 10d metric can be written as:
\begin{align}
    ds^2 
    &= \left[1 + \frac{e^{-4A_0(y)}}{c(x)}\right]^{-1/2} e^{2\Omega(x)} g_{\mu\nu}dx^{\mu}dx^{\nu}  + \left[1 + \frac{e^{-4A_0(y)}}{c(x)}\right]^{1/2}  c(x)^{1/2} g_{mn}dy^mdy^n \,,
    \label{eq:10dmetric}
\end{align}
where $e^{2\Omega(x)}$ is the Weyl-rescaling needed to go to the Einstein frame in 4d,  and we define the warp factor as 
\be\label{WF}
h\equiv1+\frac{e^{-4A_0(y)}}{c(x)} \,.
\ee 
In the limit where $c(x)\rightarrow\infty$ we identify $c(x)=\mathcal{V}^{2/3}$, with $\mathcal{V}~l_s^6$ the unwarped volume of the compact space and we take coordinates such that $V_{6} = \int d^6y\sqrt{g_6} = l_s^6$. 
We  define a warped throat as the region in which $h\gtrsim 2$, that is,  $e^{-4A_0(y)}\gtrsim c(x)$. In what follows, we will be interested in  {\em weakly warped} regions where $1\lesssim h\lesssim 2$. 

After dimensionally reducing the 10d Einstein-Hilbert term in (\ref{eq:TypeIIB}) down to 4d we identify the Weyl-rescaling parameter as
\begin{align}
    e^{2\Omega(x)} = \frac{V_{\rm w}^0}{c(x)^{3/2}\int d^6y \sqrt{g_6} ~h} = \frac{V_{\rm w}^0}{V_{\rm w}},
    && V_{\rm w} = ~c(x)^{3/2}\int d^6y \sqrt{g_6} ~h\,.
	\label{eq:Vw_definition}
\end{align}
We further choose\footnote{This choice is arbitrary and all mass scales in Planck units will be independent of the normalisation of $e^{2\Omega(x)}$. } $V_{\rm w}^0 \equiv \langle V_{\rm w} \rangle$, such that $\langle e^{2\Omega(x)}\rangle =1$. That is, the two frames are the same at the vev and the string and 4d Planck scales are related via the vev of the volume modulus. 
 In the limit of small warping (or for no warping at all)  $c(x)\gg e^{-4A_0(y)}$ and $h\approx 1$ for most of the internal space and thus  $V_{\rm w}^0 \approx \langle\mathcal{V}\rangle l_s^6$. 
 With our choice of Weyl-rescaling \eqref{eq:Vw_definition}, the definition of $M_p$ is
\begin{align}
    \frac{M_p^2}{2} = \frac{1}{2\kappa_4^2} = \frac{V_{\rm w}^0}{2\kappa^2}
	\implies
	\frac{m_s}{M_p} = \frac{g_s}{\sqrt{4\pi\V_{\rm w}^0}} \,,
	\label{eq:msMp}
\end{align}
where $ V_{\rm w}^0 =\V_{\rm w}^0 ~l_s^6$. 

The mass associated with the Kaluza-Klein towers of massive states 
for modes localised in the bulk, and thus experiencing $h \sim 1$, is given by\footnote{The factor of $2\pi$ follows by analogy with a toroidal compact space, for which $(2\pi R_{CY})^6=\langle\V\rangle l_s^6$.}
\begin{align}
	m_{KK} 
    = \frac{2\pi}{\langle\V\rangle^{1/6}}m_s \approx \frac{2\pi g_s}{\sqrt{4\pi}\V^{2/3}} M_p \,.
	\label{eq:Mkk_bulk}
\end{align}

\subsection{The warped deformed conifold}

The region of the compact space which takes the form of a (warped) deformed conifold is described by the metric \cite{candelas1990conifolds,Minasian:1999tt,Aganagic:1999fe,KS2000supergravity} 
\begin{align}
    ds_{con}^2 = \frac{\epsilon^{4/3}}{2}\mathcal{K}(\eta) 
    \Bigg( \frac{1}{3\mathcal{K}^3(\eta)} \left(d\eta^2 + (g^5)^2 \right) 
    &+ \sinh^2(\eta/2) \left( (g^1)^2 + (g^2)^2 \right) \nonumber \\
    &+ \cosh^2(\eta/2) \left( (g^3)^2 + (g^4)^2 \right)
    \Bigg)\,,
    \label{eq:deformed_conifold_metric}
\end{align}
where $\epsilon$ is a non-zero complex parameter ($\epsilon=0$ for the singular conifold),  $\mathcal{K}(\eta) = \frac{(\sinh(2\eta) - 2\eta)^{1/3}}{2^{1/3}\sinh(\eta)}$ and $g^i$ are a basis of one-forms found in \cite{candelas1990conifolds}. 

Near the tip of the conifold, the topology of the metric is $S^2 \times S^3$ with the $S^3$ of finite size and  the full 10d metric becomes \cite{KS2000supergravity}
\begin{align}
    ds_{10}^2 &= h^{-1/2} g_{\mu\nu}dx^\mu dx^{\nu} + h^{1/2}c(x)^{1/2} \left(dr_0^2 + \frac{r_0^2}{8}d\Omega_{S^2}^2 + R_\epsilon^2 d\Omega_{S^3}^2 \right),
\end{align}
where $r_0^2 = \frac{\epsilon^{4/3}}{4}\left(\frac{2}{3}\right)^{1/3}{\eta}^2$, $R_\epsilon^2 = \frac{\epsilon^{4/3}}{2}  \left(\frac{2}{3}\right)^{1/3}$, and the varying part of the warp factor is given by 
\begin{align}
    e^{-4A_0(\eta)} &=
    2^{2/3}\frac{(\alpha' g_sM)^2}{\epsilon^{8/3}} I(\eta),
    \quad\quad I(\eta)\equiv \int_{\eta}^{\infty} dx ~ \frac{x\coth(x) - 1}{\sinh^2{x}}(\sinh(2x)-2x)^{1/3}\,.
	\label{eq:warp_factor}
\end{align}
Here $M$ is the 3-form RR-flux through the $S^3$ at the tip of the throat,
\begin{align}
	\frac{1}{(2\pi)^2\alpha'}\int_{S^3} F_3
	= M\,.
\end{align}
The physical size of the $S^3$ at the tip is $R_{S^3} = h^{1/4}c(x)^{1/4}R_{\epsilon}$.  In the strongly-warped regime, this gives the well-known $R_{S^3} \sim \sqrt{\alpha' g_sM}$; the $\alpha'$-expansion is thus well under control when $g_sM\gg 1$. In contrast, in the weakly-warped scenario where $h\sim 1$ near the tip, $R_{S^3} \sim h^{1/4}c(x)^{1/4}R_{\epsilon} \sim c(x)^{1/4}\epsilon^{2/3}$, so that a controlled $\alpha'$-expansion requires instead $c(x)^{1/4}\epsilon^{2/3}/\sqrt{\alpha'}\gg 1$.

For large $\eta$, the metric (\ref{eq:deformed_conifold_metric}) approaches the singular conifold metric \cite{candelas1990conifolds,Minasian:1999tt,KS2000supergravity,KT} 
\begin{align}
	ds_{con}^2 
	&= dr_{\infty}^2 + r_{\infty}^2\left(\frac{1}{9}(g^5)^2 +\frac{1}{6}\sum_{i=1}^{4}(g^i)^2 \right),
	\quad \text{ as } \eta\to\infty\,,
	\label{eq:metric_conifold_largetau}
\end{align}
where the coordinate $r_\infty$ is defined as $r_{\infty}^2 =  \frac{3}{2^{5/3}}\epsilon^{4/3}e^{2\eta/3}$.  The warp factor becomes \cite{KT}
\begin{align}
	e^{-4A_0} \approx \frac{L^4}{r_\infty^4}\left[1 + \frac{3 g_sM}{8\pi K} + \frac{3g_s M}{2\pi K}\log\Big(\frac{r_{\infty}}{r_{UV}}\Big) \right] \,,
	\label{E:mouthwarp}
\end{align}
where we define 
\begin{align}\label{eq:L}
	L^4 = \frac{27\pi}{4}\frac{g_sMK}{(2\pi)^4}~l_s^4 \,,
\end{align}
with $K$ being the NSNS-flux number threading the 3-cycle that goes along the throat, which is dual to the $S^3$ at the tip,
\begin{align}
	\frac{1}{(2\pi)^2\alpha'}\int_{\eta\leq\eta_{\Lambda}}\int_{S^2} H_3
	= K \label{eq:H3_flux_K}\,.
\end{align}
The cut-off $\eta_\Lambda$ in this integral will be defined in the next subsection.

\subsection{Gluing the conifold to a compact CY}

Combining our discussion on the volume modulus and warped deformed conifold, we use only a finite portion of the non-compact Klebanov-Strassler solution, with the latter assumed to be glued smoothly onto a compact Calabi-Yau 3-fold.  We thus introduce a cutoff scale, $\Lambda_{UV} \sim r_{UV}$ where the conifold connects to the bulk, which is related to $\eta_\Lambda$ via
\begin{align}
    r_{UV}^2 =  \frac{3}{2^{5/3}}\epsilon^{4/3}e^{2\eta_\Lambda/3}.  
\end{align}
A natural choice for this cut-off is where  $e^{-4A_0(r)} \sim c(x)$ in the warp factor \eqref{WF}, such that $r_{UV}\sim \frac{L}{c^{1/4}}$; note that this is the minimum choice one can make for the cut-off so that the conifold is unwarped in the gluing region.  However, one may choose to extend the conifold further, thereby including an unwarped piece, before gluing onto the compact CY. 

\begin{figure}
    \centering 
    \includegraphics{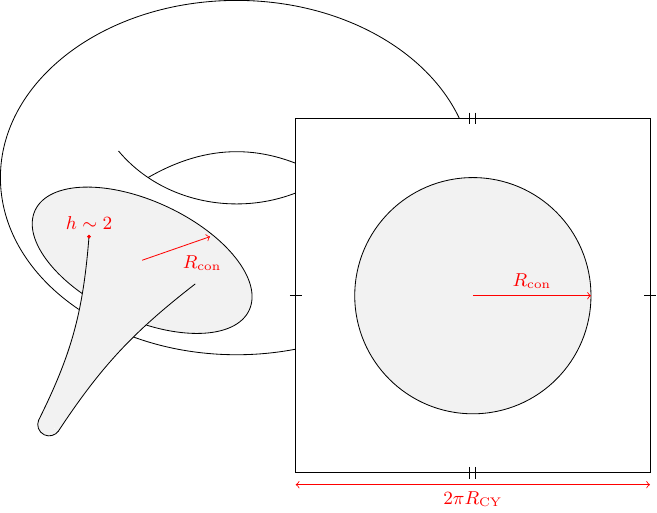}
    \caption{A cartoon of the conifold fitting into the bulk. The precise numerical factors depend on the geometry of the compact CY onto which the conifold is glued (see e.g. Appendix A of \cite{Gao:2020xqh}). If we consider an isotropic toroidal compactification of size $2\pi R_{\rm CY}$, then the conifold will fit into the bulk when $R_{con} < \pi R_{\rm CY}$. Note that one may choose to extend the conifold beyond the point where the warp factor \eqref{WF} becomes $h\sim 2$, thereby including a weakly-warped and even unwarped piece, before gluing onto the compact CY.}
    \label{fig:conifold-radius}
\end{figure}

Note that $r_{UV}$ thus defined is also related to the radial size around the top of the conifold as $R_{con} \equiv h^{1/4}(r_{UV})~c^{1/4}~r_{UV} \sim c^{1/4}~r_{UV}$.  
We need $R_{con} \lesssim \langle \mathcal{V}\rangle^{1/6} l_s$, which implies $\Lambda_{UV}\leq l_s$, in order for the conifold to `fit into' the (roughly isotropic) bulk (Fig. \ref{fig:conifold-radius}). 

Upon dimensional reduction, the deformation parameter $\epsilon$ becomes a complex structure modulus, which is part of a chiral superfield in the 4d effective action, denoted by $|S|=\epsilon^2$, with units of $(length)^3$, or the dimensionless $z=S/l_s^{3}$.  The volume modulus $c(x) = \V^{2/3}$ also falls into a chiral superfield with scalar component $\rho$ and $c = \imrho$. The radial cutoff in these coordinates is then given by
\begin{align}
	\Lambda_0^2 = \frac{3}{2^{5/3}}|z|^{2/3}e^{2\eta_{\Lambda}/3}\, ,
	\label{eq:cutoffscales_tau_Lambda}
\end{align}
where $\Lambda_0 = \Lambda_{UV}/l_s$, with $\Lambda_0\leq 1$. 

From (\ref{eq:cutoffscales_tau_Lambda}) it follows that 
\begin{align}
	K &= \frac{1}{(2\pi)^2\alpha'}\int_{\eta\leq\eta_{\Lambda}}\int_{S^2} H_3
	\approx \frac{g_s M}{2\pi}\eta_{\Lambda}
	= \frac{g_s M}{2\pi}\Big(\log\frac{\Lambda_0^3}{|z|} + \frac{3}{2}\log\frac{2^{5/3}}{3}\Big)\,,
	\label{eq:relation_K_Lambda0}
\end{align}
which therefore, neglecting contributions from the bulk, must be imposed as a consistency condition for the solution $|z|$ together with the parameters $g_s,M,K,\Lambda_0$. This is a higher-dimensional, topological derivation of the GKP solution $z \sim \Lambda_0 ~e^{\frac{2\pi K}{g_sM}}$, which follows also from the supersymmetry condition $D_zW=0$ for the superpotential derived below.  It shows that the flux number $K$ is not an independent parameter, but  is related to the parameters $g_s,M$ and $\Lambda_0$, which is nothing but a reflection of the fact that the warped conifold solution \cite{KS2000supergravity} only has a free flux number $M$, since it is a solution for constant dilaton and therefore satisfies the relation $g_s^2|F_3|^2=|H_3|^2$.

It will be useful to bear in mind the warp factor at the tip of the throat,
\begin{align}
	e^{-4A_0^{tip}} = \frac{2^{2/3}I(0)}{(2\pi)^4}\frac{(g_sM)^2}{|z|^{4/3}} \,,
	\label{eq:warp_tip}
\end{align}
and the hierarchy between the bulk (UV), where $e^{-4A_{UV}}\ll c(x)$, and the tip of the throat (IR) is
\begin{align}
	\frac{h_{IR}}{h_{UV}} = \frac{1 + \frac{e^{-4A_{0}^{tip}}}{c(x)}}{1 + \frac{e^{-4A_{UV}}}{c(x)}} \approx 1 + \frac{e^{-4A_{0}^{tip}}}{c(x)} \,,
	\label{eq:hierarchy}
\end{align}
which is large for $e^{-4A_{0}^{tip}}\gg c(x) = \V^{2/3}$.

\subsection{The low-energy effective field theory}
\label{S:KW}

The  K\"ahler potential for a type IIB flux compactification can be written as
\cite{giddings2003scales}
\newcommand{\addheight}{\parbox[c][1cm][c]{0pt}{}}
\begin{align}
    \mathcal{K}/M_p^2 =& 
    \underbrace{\addheight -2\log\left(\V_{\rm w} 
    \right)}_\text{K\"ahler moduli}
    \underbrace{\addheight -\log\left(-i(\tau-\Bar{\tau})\right)}_\text{Dilaton}
    \underbrace{\addheight -\log\left(\frac{i}{l_s^6}\int h~\Omega\wedge\Bar{\Omega}\right)}_\text{Complex structure moduli} \,.
    \label{eq:KahlerPotential1}
\end{align}
 Here and below we keep $V_{6} = l_s^6$ explicit for clarity. 
\ni The metric in the complex structure moduli space is given by (for $|z|\ll\Lambda_{0}^3$)  \cite{douglas2007warping}
\begin{align}
        G_{z\Bar{z}}
    = \frac{l_s^6}{\pi||\Omega||^2V_{6}} 
    \left(\log\frac{\Lambda_{0}^3}{|z|}
    + \frac{c' }{(2\pi)^4}\frac{1}{\V^{2/3}} \frac{(g_sM)^2}{|z|^{4/3}}
    \right)\,,
    \label{eq:appendixC_GSS}
\end{align}
where we define the constant $c'=1.18$, and  $||\Omega||^2 = \frac{1}{3!}\Omega_{mnp}\Omega^{mnp} = 8$. This metric corresponds to a contribution to the K\"ahler potential of the form
\begin{align}
    \mathcal{K}(z,\Bar{z}) =\frac{l_s^6
    }{\pi||\Omega||^2V_{6}} \left[|z|^2\left(\log\frac{\Lambda_0^3}{|z|} + 1\right) + \frac{9c'(g_sM)^2}{(2\pi)^4\V^{2/3}}|z|^{2/3}\right]\,. \label{E:K}
\end{align}
From here we see that the warping contribution to the K\"ahler potential mixes the deformation modulus $z$ and the volume modulus $\V$, in such a way that large volumes suppress the effect of the warping. Combining  the K\"ahler potentials for the two moduli we can write  (see e.g. \cite{Dudas:2019pls})
\begin{align}
    \mathcal{K} = -3\log\left(\V^{2/3} - \frac{1}{(2\pi)^4}\frac{3c'(g_sM)^2}{\pi||\Omega||^2}\frac{l_s^6}{V_{6}}|z|^{2/3}\right) + \dots \quad .
\end{align}
Notice that despite the mixing between $z$ and $\V$, the no-scale structure of the K\"ahler potential is preserved \cite{giddings2006dynamics,Blumenhagen:2019qcg}.

The superpotential  takes the well-known Gukov-Vafa-Witten form\footnote{
The normalisation of the superpotential $W$ changes depending on the way we write the volume modulus and axio-dilaton terms in the K\"ahler potential. If we write $-3\log(-i(\rho-\Bar{\rho}))$, where $\imrho = c(x) = \V^{2/3}$, there is an extra factor of $2^{3/2}$ in $W$. If the K\"ahler potential for the axio-dilaton is written as $-\log(\Im\tau)$, there is an extra factor of $2^{-1/2}$. Similarly, if the transformation between the string and Einstein frame metrics, $G_{\mu\nu}^S = e^{\frac{\phi-\langle\phi\rangle}{2}}G_{\mu\nu}^E$, does not include the vev $\langle\phi\rangle$, the factor of $g_s$ will not be present.  } \cite{Gukov:1999ya}
\begin{align}
    W/M_p^3 =  \frac{g_s^{3/2}}{\sqrt{4\pi}\cdot l_s^5} \int G_3 \wedge \Omega\,,
	\label{eq:GVW}
\end{align}
which  takes the final form \cite{GKP,douglas2007warping}
\begin{align}
    W/M_p^3 = \frac{g_s^{3/2}}{\sqrt{4\pi}}\left[W_0\,e^{i\sigma}-\frac{M}{2\pi i}z\left(\log\frac{\Lambda_0^3}{z} + 1\right) - i\frac{K}{g_s}z\right] \,.
\label{E:W}
\end{align}
Here, $W_0\,e^{\sigma}$ contains the contribution from bulk fluxes, with $\sigma$ some constant phase.
We will assume that the bulk complex structure moduli and the dilaton are stabilised by the bulk fluxes, with $\langle\tau\rangle = i g_s^{-1}$ and $W_0 e^{i\sigma}$ evaluated at the constant vevs.

In uplifting scenarios like KKLT and LVS, an $\antiD$-brane is placed at the tip of the warped throat in order to uplift the potential from an AdS minimum for the volume and conifold moduli into de Sitter. 
The potential is obtained from the brane action in the warped background
\begin{align}
    S_{\antiD} = S_{\rm DBI} + S_{\rm CS} =  -2 T_3 \int d^4x \sqrt{-\det G_{\mu\nu}} \,,
   \end{align}
where $G_{\mu\nu} = h^{-1/2}e^{2\Omega(x)}g_{\mu\nu}$ is the Einstein frame metric, giving
\begin{align}
    S_{\antiD} 
    &= \blue{-} 2T_3 \int d^4x \sqrt{-\det g_ {\mu\nu}} \left(\frac{V_{\rm w}^0}{V_{\rm w}}\right)^2h^{-1}\,.   
\end{align}
Using $T_3 = \frac{1}{(2\pi)^3g_s\alpha'^2}$, the relation between $m_s$ and $M_p$ (\ref{eq:msMp}), and $\V_{\rm w}\approx \V$, we obtain
\begin{align}
    V_{\antiD} = \left(\frac{g_s^3}{8\pi}\right)\frac{2}{\V^2}h^{-1} M_p^4\,.
    \label{eq:D3brane_potential_Mp}
\end{align}
Since the $\antiD$-brane is placed at the tip of the throat, it is common to take $e^{-4A_0(\eta_{brane})}\gtrsim c(x)$ and therefore $h\approx \frac{e^{-4A_0(\eta_{brane})}}{c(x)}$. Given that the potential is minimised for $\eta_{brane}=0$,  $V_{\antiD}$ becomes (cf. \ref{eq:warp_tip}) 
\begin{align}
    V_{\antiD} = \left(\frac{g_s^3}{8\pi}\right)\frac{(2\pi)^4}{\V^{4/3}}c''\frac{|z|^{4/3}}{(g_sM)^2} M_p^4\,.
    \label{eq:D3brane_potential_strongwarping}
\end{align}
where  $c'' = \frac{2^{1/3}}{I(0)} \approx 1.75$. However, as our interest is in the weakly-warped case, where $h \lesssim 2$
throughout the conifold, we will use the exact expression\footnote{This corrects an error in \cite{Bento:2021nbb}, where \eqref{eq:D3brane_potential_strongwarping} was used inconsistently with the weakly-warped regime.} (\ref{eq:D3brane_potential_Mp}). Note that the $\antiD$-brane can  be described in a supersymmetric way within the low energy effective supergravity theory using constrained superfields \cite{Kallosh:2014wsa,Bergshoeff:2015jxa,Kallosh:2015nia,Garcia-Etxebarria:2015lif,Dasgupta:2016prs,Vercnocke:2016fbt,Kallosh:2016aep, Aalsma:2017ulu, GarciadelMoral:2017vnz, Cribiori:2019hod}, but this will not be important for our purposes.

%% file: Sections/DeformationModulusStabilisation.tex
\section{Deformation modulus stabilisation and de Sitter vacua}
\label{sec:deformation-stab}

In this section we review the deformation modulus stabilisation in the weakly-warped regime, before including the volume modulus stabilisation by turning to the Large Volume Scenario \cite{LV1,originalLVS}, where a  new metastable de Sitter minimum was found in \cite{Bento:2021nbb}.  

Combining  the K\"ahler potential and superpotential  above, eqs.~(\ref{E:K}) and (\ref{E:W}), the resulting scalar potential is
\begin{align}
    V_{\rm KS} &= \left(\frac{g_s^3}{8\pi}\right)\frac{\pi g_s}{\V^2}
   \left(\log\frac{\Lambda_0^3}{|z|} + \frac{1}{(2\pi)^4}\frac{c'(g_sM)^2}{\V^{2/3}|z|^{4/3}}\right)^{-1}\left|\frac{M}{2\pi}\log\frac{\Lambda_0^3}{z}-\frac{K}{g_s}\right|^2 M_p^4\,.
   \label{eq:secPotential_conifoldPotential}
\end{align}
In addition to this scalar potential originating from the fluxes, we add the  contribution from a probe $\overline{\text{D3}}$-brane (\ref{eq:D3brane_potential_Mp}) at the tip of the conifold, 
\begin{align}
    V_{\antiD} = \cD\left(\frac{g_s^3}{8\pi}\right)\frac{2}{\V^2}\Bigg\{1 + \frac{1}{(2\pi)^4}\frac{2}{c''}\frac{(g_sM)^2}{\V^{2/3}|z|^{4/3}}\Bigg\}^{-1} M_p^4\,,
    \label{eq:secPotential_branePotential}
\end{align}
where $\cD=1$ in the presence of the $\overline{\text{D3}}$-brane, and zero otherwise.  The deformation modulus appears in the brane potential through the warp factor of the metric, and we see that the suppression is provided by the vev of this modulus, through $|z|^{4/3}$. In the classic KKLT scenario \cite{KKLT}, this energy suppression  ensures that the positive energy density from the probe D-brane  uplifts an otherwise AdS minimum for the volume modulus to a near Minkowki minimum, instead of dominating the potential and causing a runaway.  How much suppression is required, and hence how large the hierarchy (\ref{eq:hierarchy}) needs to be, depends on the stabilisation mechanism of the volume modulus and, in particular, on the depth of the AdS minimum prior to the uplift. 

In \cite{Bento:2021nbb} we introduced the useful parameter
\begin{align}
    \beta \equiv \frac{\V^{2/3}\log\frac{\Lambda_0^3}{\zeta}}{\frac{c'}{(2\pi)^4}\frac{(g_sM)^2}{\zeta^{4/3}}} 
	= C \, \V^{2/3}\Lambda_0^4 ~x ~e^{-\frac{4}{3}x}, \label{E:beta}
\end{align}
with $z=\zeta e^{i\theta}$, $C=\frac{(2\pi)^4}{c'(g_sM)^2}$, and $x\equiv\log\frac{\Lambda_0^3}{\zeta}$.
The parameter $\beta$ is useful because it determines the warping regime, with large values of $\beta\gg 1$ reflecting weak warping --- this can be seen both in the deformation modulus metric in (\ref{eq:secPotential_conifoldPotential}) and in the warp factor in the brane potential (\ref{eq:secPotential_branePotential}). 
Indeed from the definition (\ref{E:beta}),
\be
\beta \approx \frac{\V^{2/3}}{e^{-4A_0^{tip}}} \log\frac{\Lambda_0^3}{\zeta} 
= (h-1)^{-1}\log\frac{\Lambda_0^3}{\zeta} \,,
\ee
so that leaving the $\beta\ll 1$ regime requires us to have weak warping. Notice in particular that $\beta\propto \zeta^{4/3}\V^{2/3}$, so that the warping regime relies on the competition between small $\zeta$ and large $\V$.
Defining also the constants
\begin{align}
    \varepsilon = \frac{g_s M}{2\pi K}\,, && 
    \delta_1 = \frac{g_s^3}{8}\times\frac{K^2}{g_s} \,, &&
    \delta_2 = \frac{g_s^3}{8\pi} \times c'' \frac{c'}{\delta_1} = \frac{1}{\pi}\times c'' c' \frac{g_s}{K^2} \,, &&
    \delta_3 \equiv \frac{(2\pi)^4}{2}\frac{c''}{(g_s M)^2} \,,
\end{align}
the potentials  (\ref{eq:secPotential_conifoldPotential}) and (\ref{eq:secPotential_branePotential}) become (in Planck units,  $M_p=1$)
\begin{align}
    V 
    &= \delta_1 C \Lambda_0^4\frac{e^{-\frac{4}{3}x}}{\V^{4/3}}  
    \left[(1+\beta)^{-1}(1-\varepsilon x)^2 
    + \cD\delta_2\Big(1 + \delta_3 \V^{2/3}\Lambda_0^4e^{-\frac{4}{3}x}\Big)^{-1}\right] \,,
    \label{eq:VKSpVD3}
\end{align}
with the axion already fixed at $\langle\theta\rangle=0$ . 

Since the weakly-warped regime where the warping term in (\ref{E:K}) is subdominant\footnote{To be precise, we consider a ``weakly-but-still-warped scenario'', by which we mean that the $c'$ term in the deformation modulus metric---the strong warping correction to GKP by DST \cite{douglas2007warping}---provides an important subleading correction to the scalar potential (\ref{eq:scalarpotential}), even though the interplay of all the ingredients is such that the warping is small.
Starting with a conifold provides us with an explicit background metric onto which we can add a probe anti-D3 brane. However, to stabilise the conifold modulus, we need fluxes, which then source a warped throat.} corresponds to $\beta\gg 1$, the potential is approximately
\begin{align}
    V &\approx \frac{\delta_1}{\V^2}\frac{1}{x}\left[\Big(1-\frac{1}{\beta}\Big)\Big(1-\varepsilon x\Big)^2 + \beta ~\cD\delta_2\Big(1 + \delta_3 \V^{2/3}\Lambda_0^4e^{-\frac{4}{3}x}\Big)^{-1}\right] \,.
	\label{eq:scalarpotential}
\end{align}
The minimum can be found perturbatively in $1/\beta$, 
\begin{equation}
    x_{min} = x_0 + \frac{x_1}{\beta} + \mathcal{O}\Big(\frac{1}{\beta^2}\Big) \,,
\end{equation}
plugging $x_{min}$ into $V'$ and solving at each order in $1/\beta$, which gives
\begin{equation}
    x_{min} = \frac{1}{\varepsilon} + \cD\frac{2 \delta_2 C^2}{3\delta_3^{\phantom{3}2}\varepsilon^4}\frac{1}{\beta}\Bigg|_{x=1/\varepsilon} 
    + \mathcal{O}\Big(\frac{1}{\beta^2}\Big) \,.
\end{equation}
At leading order we obtain the GKP solution, $x_0 = 1/\varepsilon \Rightarrow \zeta_0 = \Lambda_0^3 \exp\{-\frac{2\pi K}{g_sM}\} \equiv \zGKP$, while 
the $1/\beta$ corrections depend explicitly on $c_{\rm D3}$, reflecting the fact that it is a correction due to the presence of the brane. This gives for $\zeta_{min}$ 
\begin{equation}\label{eq:zmin}
    \zeta_{min} \approx \zGKP \cdot\exp\left\{
    - \cD\frac{4 K }{3\pi^2 c' M \zGKP^{4/3} \V^{2/3}} \right\} \,,
\end{equation}
corresponding to a small shift of the GKP solution towards smaller values of $z$. Note the presence of the combination $\zGKP^{4/3}\V^{2/3}$ in the correction, which reminds us of the $1/\beta$ suppression and how it relies on the competition between small $\zeta$ and large $\V$. We always find a solution provided $\beta$ is large enough so that the expansion in $1/\beta$ is valid, showing that in the weakly-warped limit there is no uplifting runaway bound on the fluxes and the anti-brane does not destabilise the conifold modulus (cf. \cite{upliftingrunaways2019}). Moreover, since the solution is always only a small perturbation away from $\zGKP$, we do not expect it to be affected by the discussion of \cite{Lust:2022xoq}, which argues that higher-dimensional constraint equations become important when studying the 4d potential far away from the GKP minimum\footnote{In contrast, the existence and position of the maximum that we discuss below crucially depends on the off-shell $\zeta$-dependence of the potential and therefore our discussion of the maximum will no longer apply if the potential behaves differently away from the minimum, as suggested by \cite{Lust:2022xoq}. The solution we are interested in will not be affected by a different off-shell potential, since it corresponds to the minimum, which is only a small perturbation away from the GKP solution.}.

\begin{figure}
    \centering
    \includegraphics[width=0.65\linewidth]{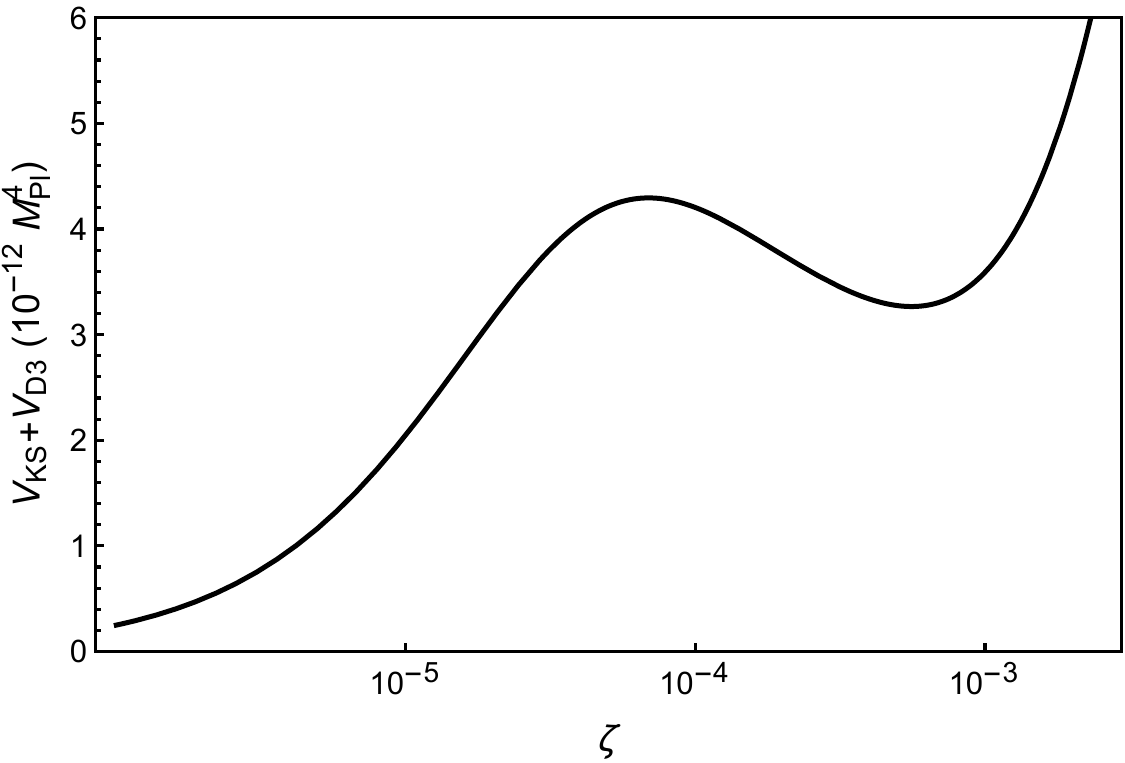}
    \caption{Potential (\ref{eq:VKSpVD3}) in the regime $\beta\gg 1$ (with the $\zeta$-axis presented in log scale), with the choice of parameters $\Lambda_0=0.43,~g_s=0.17,M=16,K=2,~\V=10^4$ and given $||\Omega||^2 = 8,~c'=1.18,~c''=1.75$; for this parameter set $\zeta_{min}\approx 5.57\times 10^{-4}$ and $\zeta_{max}\approx 6.89\times 10^{-5}$. For comparison, $\zGKP\approx 7.83\times 10^{-4}$. 
    } 
    \label{fig:Vlargebeta}
\end{figure}

Since $V\to 0$ as $\zeta\to 0$ \footnote{Note that this is only true because the warping correction to the deformation modulus metric is still present. The $\Op{1/\beta}$ correction itself relies on the warping being \emph{small but non-vanishing}.} and there is a minimum, there must be a maximum at some $0<\zeta_{max}<\zeta_{min}$. If $\zeta_{max}$ is sufficiently smaller than $\zeta_{min}$, we will have $\varepsilon x_{max} \gg 1$. Taking this limit  
and in terms of the variable $y \equiv h-1 = \frac{2}{(2\pi)^4}\frac{(g_s M)^2}{c'' \V^{2/3}\zeta^{4/3}}\neq 0$, one can show that $V'$ is proportional to
\be 
    V' \propto 
     -y^3
     +\left(\frac{3}{2c'c''} - 2\right) y^2
     +\left(\frac{3}{c'c''}-\frac{16\pi \cD}{c'c''(g_s M^2)} - 1\right) y
     +\frac{3}{2c'c''} \,.
\ee 
The sign of the discriminant of this cubic equation is only a function of $(g_sM^2)$ and always negative --- there can only be one real solution. On the other hand, the variable $y = h - 1 < 1$ and the coefficients are all $\mathcal{O}(1)$ or more. Hence, neglecting the cubic term, we find the approximate solution for the maximum,
\begin{align}
    y =&~ \frac{ (3-c' c'') (g_s M^2)- 16\pi \cD}{(g_s M^2)  (4 c c''-3)} \nonumber \\
    &+ \frac{\sqrt{c'  c'' (c c''+6)(g_s M^2)^2 - 32\pi \cD(g_s M^2) (3 - c' c'') + 64 (2\pi)^2\cD^2}}{(g_s M^2)  (4 c' c''-3)}
    \,.
    \label{eq:ymax}
\end{align} 
In Fig.\ref{fig:Vlargebeta} we plot the potential (\ref{eq:VKSpVD3}) in the regime $\beta\gg 1$ for a specific set of parameters, which shows both the minimum near the GKP solution and a maximum at $\zeta_{max}$. Note also that the value of $\zeta_{min}$ in the weakly-warped case should not be too small in order for $\beta$ (\ref{E:beta}) to be large. 

Since $\beta\gg 1$ is also associated with large volumes, it is natural to use the Large Volume Scenario \cite{LV1,originalLVS} to explicitly stabilise the volume modulus.  This relies on balancing leading perturbative and non-perturbative contributions to the 4d low-energy effective field theory. Several further corrections to the scalar potential may arise from warping, curvature and string-loop effects, which may affect the stabilisation studied in \cite{Bento:2021nbb}. In the Section \ref{sec:ww-LVS-dS} we review the construction in \cite{Bento:2021nbb} and in Section \ref{sec:dangerous-corrections} we will analyse the effect of sub-leading corrections on this leading solution. 

%% file: Sections/LVS.tex
\section{Weakly-warped LVS de Sitter vacua}
\label{sec:ww-LVS-dS}

In this section we  review the full moduli stabilisation and dS solution in the weakly-warped regime $(\beta\gg 1)$, including the coupled system of K\"ahler and complex structure moduli, by embedding the above system into the Large Volume Scenario. The bulk K\"ahler moduli field content is therefore extended to include the two moduli of a ``Swiss cheese'' Calabi-Yau. 

In the Large Volume Scenario \cite{originalLVS}, the K\"ahler moduli are stabilised by balancing different quantum corrections against each other. It relies on both perturbative $\alpha'$ corrections to the K\"ahler potential\footnote{Note that the form of the correction term depends on the convention used for the change between string frame and Einstein frame \cite{Becker:2002nn}. In our conventions, it reads (in terms of the Einstein frame volume $\V$)
\be
-2\log\Big[\mathcal{V}+\frac{\xi}{2} e^{-\frac{3}{2}(\phi - \langle\phi\rangle)}\Big] 
= -2\log\Big[\mathcal{V}+\frac{\xi}{2}\Big] \,,
\ee
once the dilaton $\phi$ is stabilised at $\langle e^{\phi}\rangle = g_s$. This is in contrast with \cite{LVSdS:2010.15903,Junghans:2022exo}, in which the change of frames does not include the dilaton vev. Notice that also the definition of $\xi$ differs by a factor of $2$.} and non-perturbative corrections to the superpotential,
\begin{align}
    \mathcal{K}/M_p^2 &= -2\log\Big[\mathcal{V}+\frac{\xi}{2}\Big]\,, \\
    W &= W^{cs} + \sum_i A_i e^{-\frac{a_i}{g_s} T_i}\,,
\end{align}
where $\xi = -\frac{\chi(X_6)\zeta(3)}{2(2\pi)^3}$, $\chi(X_6)$ being the Euler characteristic of the Calabi-Yau 3-fold $X_6$ that describes the internal space with $\zeta(3)\approx 1.202$, $W^{cs}$ is the flux superpotential responsible for stabilising the complex structure moduli and the dilaton, $i$ runs over the K\"ahler moduli and the real part of $T_i = \tau_i + i\theta_i$ are the volumes of internal four-cycles, being therefore related to the volume of the internal space, while the axions $\theta_i$ correspond to deformations of the RR-form $C_4$. The superpotential for $T_i$ arises from non-perturbative effects --- it can be generated by gaugino condensation on a stack of D7-branes ($a_i=\frac{2\pi}{N}$, where $N$ is the rank of the world-volume gauge group) or by Euclidean D3-brane instantons ($a_i=2\pi$); the coefficients $A_i$ may generically depend on complex structure moduli, the dilaton, and brane moduli that have been integrated out.
As in \cite{LVSdS:2010.15903,Bento:2021nbb}, we consider a CY 3-fold with $h^{1,1}=2$ and a ``Swiss cheese'' form for the volume 
\begin{align}
    \mathcal{V} = \tau_b^{3/2} - \kappa_s \tau_s^{3/2}\,,
\end{align}
 where only the leading non-perturbative effect is considered \cite{LV1,originalLVS,LVSdS:2010.15903,Bento:2021nbb},
\begin{align}
    W = W^{cs} + Ae^{-\frac{a}{g_s}T_s}\,.
\end{align}
Therefore, we consider the following K\"ahler and super potentials 
\begin{align}
    \mathcal{K}/M_p^2 =& -2\log\left[\mathcal{V} + \frac{\xi}{2}\right] 
    -\log\left(-i(\tau-\Bar{\tau})\right) 
    -\log\left(||\Omega||^2\right)\nonumber \\
    & + \frac{1}{\pi||\Omega||^2}\left(|z|^2\left(\log\frac{\Lambda_0^3}{|z|} + 1\right)
    +\frac{1}{(2\pi)^4}\frac{9c'(g_sM)^2}{\mathcal{V}^{2/3}} |z|^{2/3}\right) \,, \\
    W/M_p^3 =&~ \frac{g_s^{3/2}}{\sqrt{4\pi}}\left(W_0\,e^{i\sigma} + \left[-\frac{M}{2\pi i}z\left(\log\frac{\Lambda_0^3}{z} + 1\right) - i \frac{K}{g_s} z \right] 
    + Ae^{-\frac{a}{g_s}T_s}\right)\,,
	\label{eq:LVS_Kahler_w}
\end{align}
where $z=\zeta e^{i\theta}$. From these definitions we have that the gravitino mass is\footnote{Note that the exact factors of $g_s$ in the gravitino mass are convention dependent when expressed in terms of the Einstein frame volume $\V$ (see footnote \ref{F:vev_shift}) \cite{ValeixoBento:2023afn}.}
\begin{align}
	m_{3/2} = e^{\mathcal{K}/2}|W| \approx \frac{g_s^2W_0}{\sqrt{8\pi}||\Omega||\V} M_p\,. 
	\label{eq:mgravitino_bulk}
\end{align}

Computing the scalar potential $V$ in the limit $\mathcal{V}\gg 1$ and $\zeta\ll 1$ gives\footnote{One must be careful with the term $K^{z\bar{z}}(D_zW)(D_{\bar{z}}\bar{W})$, since there is a competition between these two limits, whose result depends on the $\beta$-regime we are working in. Usually, it is assumed that the warp factor completely dominates the bracket ($\beta \ll 1$) in the second line, which is equivalent to taking the limit with the constraint $\zeta^{4/3}\mathcal{V}^{2/3}\ll 1$. By introducing $\beta$ before expanding the potential around $1/\V$, one may take the two regimes into account. 
}
\begin{align}
    V &= \frac{g_s^4}{8\pi||\Omega||^2}
    \Bigg(\frac{8a^2A^2\sqrt{\tau_s}e^{-2\frac{a}{g_s}\tau_s}}{3\kappa_s g_s^2\V}
    + \frac{4 aA\tau_se^{-\frac{a}{g_s}\tau_s}}{g_s\V^2}W_0\cos(\frac{a}{g_s}\theta_s + \sigma) 
    +\frac{3 W_0^2}{4\V^3}\xi \\
    &~~ + \frac{\pi ||\Omega||^2}{\V^{2}}\Big(\log\frac{\Lambda_0^3}{\zeta} + \frac{1}{(2\pi)^4}\frac{c'(g_sM)^2}{\V^{2/3}\zeta^{4/3}}\Big)^{-1}\Bigg[\frac{M^2}{(2\pi)^2}\theta^2 + \left( \frac{M}{2\pi}\log\frac{\Lambda_0^3}{\zeta} - \frac{K}{g_s}\right)^2\Bigg] \Bigg)\,. \nonumber
\end{align}
Notice that the $T_b$ axion, $\theta_b$, remains a flat direction at leading order and would be stabilised by subleading non-perturbative effects. Looking at $\partial_\theta V= \partial_{\theta_s} V= 0$, we find the solutions for the remaining axions
\begin{align}
    \langle\theta\rangle = 0 \,,&&
    \langle\theta_{s}\rangle = \frac{n\pi - \sigma}{a}g_s, \quad n\in\Z \,,
	\label{eq:axion_vevs}
\end{align}
and choose $n=1$, such that $\cos(\frac{a}{g_s}\theta_s + \sigma) = -1$. By inspecting the Hessian matrix in the axion directions, $\partial_j\partial_\theta V$ and $\partial_j\partial_{\theta_s} V$, where $j$ runs through all fields, we conclude that these completely decouple from the other moduli and therefore we can 
fix the axions to their minima and then analyse the 3-field system: $(\V,\tau_s,\zeta)$. In particular, the axion masses are always positive, making these stable directions. The potential then becomes
\begin{align}
    V =& \frac{g_s^4}{8\pi||\Omega||^2}
    \Bigg(\frac{8a^2A^2\sqrt{\tau_s}e^{-2\frac{a}{g_s}\tau_s}}{3\kappa_s g_s^2\V}
    - \frac{4 aA W_0 \tau_se^{-\frac{a}{g_s}\tau_s}}{g_s\V^2}
    +\frac{3W_0^2}{4\V^3}\xi \nonumber \\
    & + \frac{\pi ||\Omega||^2}{c'}\frac{(2\pi)^4}{\V^{4/3}}\frac{\zeta^{4/3}}{(g_sM)^2}\Big(1+\beta\Big)^{-1}\left( \frac{M}{2\pi}\log\frac{\Lambda_0^3}{\zeta} - \frac{K}{g_s}\right)^2 \Bigg) \nonumber \\ 
    &+ \cD \left(\frac{g_s^3}{8\pi}\right)\frac{2}{\V^2}\Bigg\{1 + \frac{1}{(2\pi)^4}\frac{2}{c''}\frac{(g_sM)^2}{\V^{2/3}\zeta^{4/3}}\Bigg\}^{-1}\,,
    \label{eq:LVS_full_potential}
\end{align}
where we introduced our variable $\beta$, defined in (\ref{E:beta}), and the brane potential \eqref{eq:secPotential_branePotential}.
\noindent It turns out that the solution for $\tau_b$ does not depend on the choice of $\beta$ regime, giving
\begin{align}
    \V \approx \tau_b^{3/2} = \frac{3W_0 g_s \kappa_s\sqrt{\tau_s}e^{\frac{a}{g_s}\tau_s}}{aA}\frac{a\tau_s-g_s}{4a\tau_s-g_s}
    \label{eq:solV}
\end{align}
in both cases. However, we see that the solution is given in terms of $\tau_s$, which means that there is an implicit dependence on the choice for $\beta$ hiding in the solution for $\tau_s$. In turn, both $\zeta$ and $\tau_s$  have different solutions depending on the regime of $\beta$ that we look at. 

In the weakly-warped regime $(\beta\gg 1)$, the potential becomes\footnote{Note that the warp factor in $V_{\antiD}$ can be written as
\begin{equation}
    h = 1 + \frac{2}{c'c''}\frac{\log\frac{\Lambda_0^3}{\zeta}}{\beta} \,.
\end{equation}
Here we assume that $\beta\gg\log\frac{\Lambda_0^3}{\zeta}$ and expand $h$ accordingly.}
\begin{align}
    V &\approx \frac{g_s^4}{8\pi||\Omega||^2}\Bigg\{
        \frac{8a^2A^2\sqrt{\tau_s}e^{-2\frac{a}{g_s}\tau_s}}{3\kappa_s g_s^2\V}
        - \frac{4 aA W_0 \tau_se^{-\frac{a}{g_s}\tau_s}}{g_s\V^2}
        +\frac{3 W_0^2}{4\V^3}\xi \nonumber \\
    &+ \frac{\pi||\Omega||^2}{g_s}\frac{1}{\V^{2}\log^2\frac{\Lambda_0^3}{\zeta}}
    \Bigg(\log\frac{\Lambda_0^3}{\zeta}-\frac{1}{(2\pi)^4}\frac{c'(g_sM)^2}{\V^{2/3}\zeta^{4/3}}\Bigg)\left(\frac{M}{2\pi}\log\frac{\Lambda_0^3}{\zeta} - \frac{K}{g_s}\right)^2 \nonumber \\
   &+ \cD\frac{||\Omega||^2}{g_s}\frac{2}{\V^2}\Bigg(1 - \frac{2}{c'c''}\frac{1}{(2\pi)^4}\frac{c'(g_sM)^2}{\V^{2/3}\zeta^{4/3}}\Bigg) \Bigg\} \,, \label{E:largebetaV}
\end{align}
which has a minimum at
\begin{align}
    \V &\approx \tau_b^{3/2} = \frac{3(a\tau_s-g_s)}{4a\tau_s-g_s}\cdot \frac{W_0 g_s \kappa_s\sqrt{\tau_s}}{aA}
    \cdot e^{\frac{a}{g_s}\tau_s} \,,
    \label{eq:solV} \\ 
    \tau_s^{3/2} &\approx \frac{\xi}{2 \kappa_s} 
    + \frac{g_s}{3a}
    + \frac{8||\Omega||^2\V}{9g_s\kappa_s W_0^2} + \mathcal{O}\Big(\frac{1}{\beta}\Big) \,, \label{eq:soltaus_largebeta}\\
    \zeta_{min} &\approx \zGKP \cdot\exp\left\{
    - \cD\frac{4 K }{3\pi^2 c' M \zGKP^{4/3} \V^{2/3}} \right\} \label{eq:solz_largebeta} \,.
\end{align}

In order for the third term in \eqref{eq:soltaus_largebeta} to be a correction to the leading order solution $(\tau_s^{(0)})^{3/2} = \xi/(2\kappa_s)$ in a consistent way, we must have 
\begin{equation}\label{eq:W0_condition}
    \frac{8||\Omega||^2\V}{9g_s\kappa_s W_0^2} \ll \frac{\xi}{2 \kappa_s}  \,.
\end{equation}
One can also see this by noticing that the solution for $\tau_s$ is implicit, since there is a dependence on $\tau_s$ in the third term through $\V$. If (\ref{eq:soltaus_largebeta}) is a solution, the function 
\begin{equation}
    F(\tau_s) = -\tau_s^{3/2} + \frac{\xi}{2 \kappa_s} + \frac{2||\Omega||^2}{3 a A W_0}\sqrt{\tau_s}~e^{\frac{a}{g_s}\tau_s} \,,
    \label{eq:Ftaus}
\end{equation}
where we have substituted $\V$ by its solution in the limit $a\tau_s\gg g_s$, must have a root. In particular, the exponential term cannot be too large, since it has to balance against the $\tau_s^{3/2}$, which suggests that the combination $AW_0$ must be exponentially large. We can make a better estimate by noting that $F(\tau_s)$ has a minimum, which must be non-positive for a root to exist. Again in the limit $a\tau_s\gg g_s$, this leads to the condition 
\begin{equation}
    A W_0 > \frac{4||\Omega||^2}{9g_s}\exp\left\{\frac{a}{g_s}\Big(\frac{\xi}{2 \kappa_s}\Big)^{2/3}\right\} \,,
    \label{eq:AW0_bound}
\end{equation}
which indeed corresponds to having $A W_0$ exponentially large. Note that this condition is stronger than \eqref{eq:W0_condition}, which can be seen by rewriting \eqref{eq:AW0_bound} in terms of the volume modulus vev using \eqref{eq:solV}, 
\begin{equation}\label{eq:W0V-bound}
    \frac{8||\Omega||^2\V}{9g_s\kappa_s W_0^2} < \frac{3g_s}{2a}\cdot \left(\frac{\xi}{2\kappa_s}\right)^{1/3} \,.
\end{equation}

\begin{figure}
    \centering
    \includegraphics[width=0.65\linewidth]{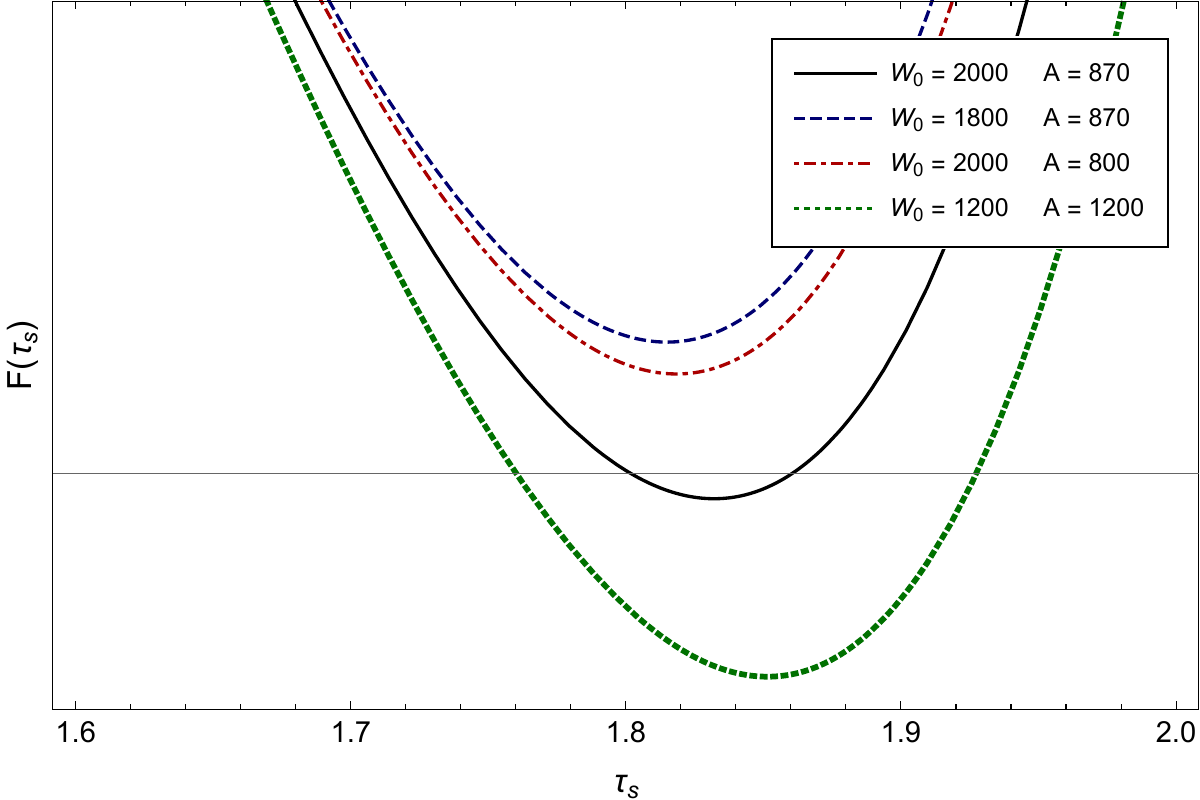}
    \caption{The plot shows (\ref{eq:Ftaus}) for different choices of the parameters $W_0$ and $A$, with all other parameters fixed. Critical points for the scalar potential (\ref{E:largebetaV}) exist when $F(\tau_s)=0$.  We see that the product $AW_0$ needs to be large enough for a solution to exist (\ref{eq:AW0_bound}).}
    \label{fig:condition_for_taus_solution}
\end{figure}

At this solution the vacuum energy is given by 
\begin{align}\label{eq:Vmin}
    V_{min} = \frac{g_s^4}{8\pi||\Omega||^2}\cdot \frac{3g_s \kappa_s W_0^2 \sqrt{\tau_s}}{4a\V^3} \Big(-1 + \alpha + \Op{g_s} + \Op{1/\beta}\Big) \,,
\end{align}
where we define the uplift parameter $\alpha$ as\footnote{Note that although the definition of $\alpha$ differs from the one in \cite{Junghans:2022exo}, it still encodes the brane uplift and its backreaction on the solution just as it did in the strongly-warped regime. It is here simply adapted to the weakly-warped regime, making explicit the need for large $W_0$ in order to suppress the effects of the brane. Note also that the $\tau_s$ in this expression corresponds to the leading term in the solution, i.e. $\tau_s^{(0)} = \frac{\xi^{2/3}}{(2\kappa_s)^{2/3}}$.} 
\begin{align}\label{eq:alpha}
    \alpha = \cD~\frac{8a||\Omega||^2}{9 \kappa_s \sqrt{\tau_s}}\frac{\V}{g_s^2 W_0^2} \,.
\end{align}
Therefore a dS vacuum requires $\alpha > 1$.

\newcommand{\Nflux}{N_\text{flux}}
Due to the exponentially large $AW_0$ required for a solution in the weakly-warped regime (which implies at least a somewhat large $W_0$ in order to keep the volume exponentially large, cf. \eqref{eq:solV}), it is important to recall a relation between the flux superpotential $W_0$ and the flux contribution to the D3 tadpole \cite{Denef:2004ze},\footnote{We thank Arthur Hebecker, Simon Schreyer and Gerben Venken for pointing this out to us.}
\begin{equation}
    \Nflux \geq \frac{g_s W_0^2}{2||\Omega||^2} \,, 
    \label{eq:DDbound}
\end{equation}
which pushes the flux contribution to the tadpole towards larger values, consequently requiring compact spaces whose topology is capable of cancelling the tadpole (cf. \cite{Gao:2022fdi,Junghans:2022kxg}). The constraint \eqref{eq:DDbound} can be derived as follows\footnote{We thank Erik Plauschinn, and also Simon Schreyer, for explaining this to us.} \cite{Erik}.  The background ISD condition, $\star_6 G_3 = i G_3$, implies that $F_3 = s(H_3-\star H_3)$, where we write $s = \Re \tau = 1/g_s$.  Expanding $H_3$ as $H_3 = h^0 \Omega + h^i \chi_i +  h^{\bar{\jmath}} \bar{\chi}_{\bar{\jmath}}+ \bar{h}^{\bar{0}} \bar{\Omega}$, it follows that $\Nflux \equiv \frac{1}{l_s^4}\int F_3 \wedge H_3$ can be written as\footnote{Recall that $\star\Omega = - i\Omega$, $\star\bar{\Omega}=i\bar{\Omega}$, $\star\chi_i = i\chi_i$, $\bar{\chi}_{\bar{\jmath}} = -i\bar{\chi}_{\bar{\jmath}}$ and $\int\Omega\wedge\bar{\Omega} = -i||\Omega||^2 l_s^6$.} $\Nflux = 2s (h^0 \bar{h}^{\bar{0}} + h^i G_{i\bar{\jmath}} h^{\bar{\jmath}}) ~\frac{i}{l_s^4}\int \Omega \wedge \bar{\Omega}$.  Since the K\"ahler metric is always positive, this implies that $\Nflux \geq 2s h^0 \bar{h}^{\bar{0}} ~\frac{i}{l_s^4}\int \Omega \wedge \bar{\Omega}$. 
At the same time, the F-term condition $F_\tau = 0$ implies that $\bar{f}^{\bar{0}}-\bar{\tau}\bar{h}^{\bar{0}}=0$ -- where $F_3$ has been expanded similarly to $H_3$ -- so that $W_0 \equiv \frac{1}{l_s^5}\int G_3 \wedge \Omega = 2s \bar{h}^{\bar{0}} \frac{i}{l_s^5}\int \Omega \wedge \bar{\Omega}$.  Eq. (\ref{eq:DDbound}) immediately follows.

\vskip 1em
\subsection*{Example}

We choose a set of parameters which guarantees both $V_{crit}>0$ and $m_3^2>0$, and thus a dS minimum. We have parameters associated with the fluxes for the remaining complex structure moduli and the axio-dilaton $(W_0,g_s)$, with the conifold $(M,K,\Lambda_0)$, and with the CY and K\"ahler moduli $(\kappa_s,\xi,a,A)$, where $\xi = -\frac{\chi \zeta(3)}{2(2\pi)^3}$. The fixed parameters in the potential are $c' = 1.18$, $c'' = 1.75$, $||\Omega||^2 = 8$.

In Table \ref{tb:parlargebeta} we present a set of parameters that serve to provide a concrete example of the stabilisation mechanism presented in this subsection. As we have seen, we need large values of $AW_0$. It may be possible to accomodate these through the specific values of the complex structure moduli that have been stabilised and integrated out. For example, we will see in the next section how one-loop corrections to the gauge kinetic function on wrapped D7-branes effectively increase the value of $A$. From the instanton point of view, this corresponds to a one-loop corrected instanton action $W_{np} \sim e^{-S_{\rm inst}}$, with $S_{\rm inst}\sim \frac{a}{g_s}\tau_s - \log A \approx 4.3 > 1$ for our numerical example, which keeps the instanton expansion under control. The solution is sourced by flux numbers $MK=32$, which straightforwardly allows tadpole cancellation using only orientifold planes within 4d perturbative type IIB. However, the large $W_0$ and condition \eqref{eq:DDbound} bring back the tadpole problem, needing constructions that allow for tadpole cancellation when $\Nflux > 42\,500$. 

\begin{table}[h!]
\centering
 \begin{tabular}{| c | c | c | c | c | c | c | c | c | c | } 
 \hline
 \rowcolor{black!10!white!90}  $W_0$ & $\sigma$ & $g_s$ & $M$ & $K$ & $\Lambda_0$ & $\kappa_s$ & $\chi$ & $a$ & $A$ \\  
 \hline\hline
$2000$ & $0$ & $0.17$ & $16$ & $2$ & $0.43$ & $\frac{\sqrt{2}}{9}$ & $-280$ & $\frac{\pi}{3}$ & $870$ \\ 
 \hline
 \end{tabular}
\caption{Choice of  parameters for the potential (\ref{eq:LVS_full_potential}), with $\beta\gg 1$.
}
\label{tb:parlargebeta}
\end{table}

For the parameter set in Table \ref{tb:parlargebeta}, we determine the critical points of the full potential (\ref{eq:LVS_full_potential}), summarised in Table \ref{tb:sollargebeta}. While one critical point is a minimum, the second is a saddle point with one unstable direction. Note that for both cases we indeed find $\beta > 1$ ($\beta_{min}\approx 7$ and $\beta_{saddle} \approx 15$).
\begin{table}[h!]
\centering
	 \begin{tabular}{| c | c | c | c | c | c | c |} 
		\hline
		\rowcolor{black!10!white!90} 
        $\tau_s$ & $\tau_b$ & $\zeta$ 
        & $V_{crit}$
        & $m_1^2\sim m_{\zeta}^2$ & $m_2^2\sim m_{\tau_s}^2$ & $m_3^2$ \\  
		\hline\hline
		1.80 & 239 & $4.17\times 10^{-4}$ 
        & $1.70\times 10^{-13}$
		& $9.23\times 10^{-5}$ 
        & $4.87\times 10^{-4}$ 
        & $4.32\times 10^{-11}$   \\ \hline
	 \end{tabular}
     \vskip 1em
     \resizebox{\columnwidth}{!}{%
     \begin{tabular}{| c | c | c | c | c | c | c | c |} 
		\hline
		\rowcolor{black!10!white!90} $\V$ & $M_s$ & $m_{KK}$ &  $m_{3/2}$ & $M_s^{\rm w}$ & $m_{KK}^{\rm w}$ & $m_{3/2}^{\rm w}$ \\  
		\hline\hline
        $3684$ 
        & $4.96\times 10^{-3}$ 
        & $1.26\times 10^{-3}$ 
        & $1.11\times 10^{-3}$ 
        & $4.33\times 10^{-3}$ 
        & $1.10\times 10^{-3}$
        & $0.96 \times 10^{-3}$ \\ \hline
	 \end{tabular}
     }
	\caption{Solution and masses for the fields $(\tau_s,\tau_b,\zeta)$ for the parameter set in Table \ref{tb:parlargebeta}, and corresponding physical scales (in units of $M_p$). The scales $M_s^{\rm w}$, $m_{KK}^{\rm w}$ and $m_{3/2}^{\rm w}$ are warped down with respect to their bulk counterparts by the factor $h_{tip}^{-1/4} = h^{-1/4}_{IR}$ (see equation \eqref{eq:hierarchy}); in the present weakly-warped case the distinction is not significant. Since all mass-squareds are positive, the solution is metastable.  
    For comparison, $\zGKP\approx 7.83\times 10^{-4}$.}
    \label{tb:sollargebeta}
\end{table}

From Table \ref{tb:sollargebeta} we see that the cutoff for the EFT is $m_{KK}\sim m_{KK}^{\rm w}$,  which reflects the fact that this solution has weak warping.
Note that the moduli masses are much lighter than the KK scale, so that the KK modes can be safely integrated out. 
At the same time, $\frac{m_{3/2}}{m_{KK}}\approx 0.88$, which is at the limit of a consistent 4d supergravity description (see also \cite{Cicoli:2013swa}).
Also $\Lambda_0 < 1$, so that the conifold fits into the bulk as in the strongly-warped case \cite{Carta_2019,Gao:2020xqh}.  
Similarly to the usual LVS, this solution avoids the bulk singularity problem \cite{Gao:2020xqh,Carta:2021lqg} due to the hierarchy $\tau_s\ll \tau_b\sim\V^{2/3}$.
 
\begin{figure}[ht]
    \centering
    \includegraphics[width=0.65\linewidth]{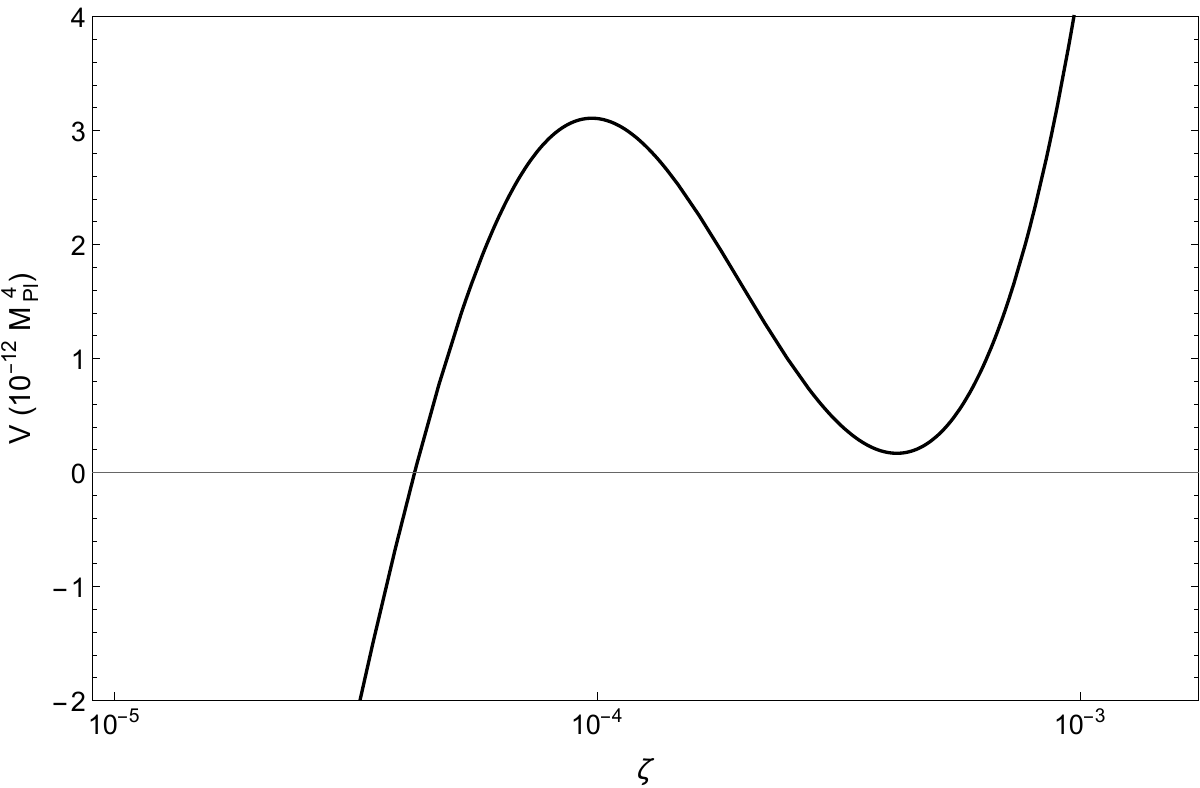}
    \caption{Plot of the potential (\ref{eq:LVS_full_potential}) in the $\zeta$ direction (the $\zeta$-axis is presented in log scale), for the parameter set in Table \ref{tb:parlargebeta}, with $(\tau_s,\tau_b)$ fixed to the values given in Table \ref{tb:sollargebeta}.  
    For comparison, $\zGKP\approx 7.83\times 10^{-4}$.}
 \label{fig:minimum_slices_largebeta}
\end{figure}

\begin{figure}[]
  	\centering
  	\includegraphics[width=0.8\linewidth]{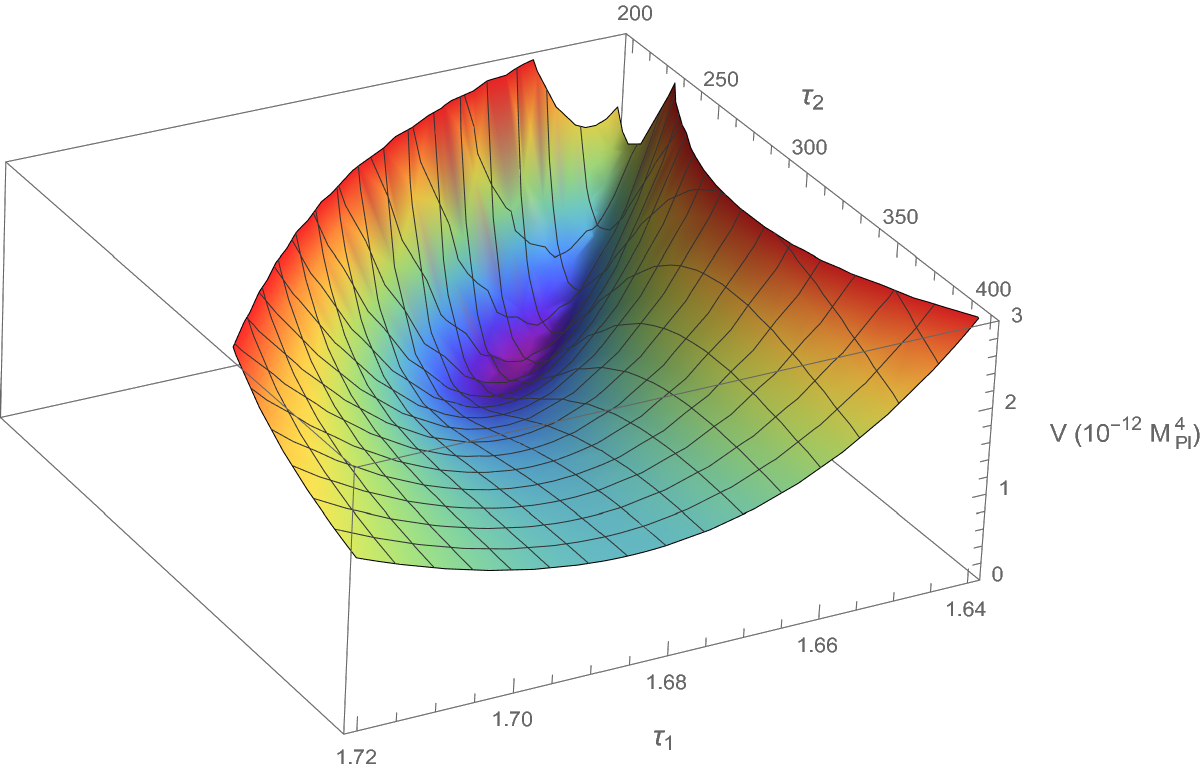}
	\caption{Plot of the potential (\ref{eq:LVS_full_potential}) in the $(\tau_1,\tau_2)$ plane (a rotation of the $(\tau_b,\tau_s)$ plane aligned with the eigenvectors of the Hessian matrix), for the parameters in Table \ref{tb:parlargebeta} (in Planck units).}
 \label{fig:minimum_plot3d_largebeta}
\end{figure}

In Figs. \ref{fig:minimum_slices_largebeta}--\ref{fig:minimum_plot3d_largebeta} we show the plots of the potential in each direction and in the plane $(\tau_s,\tau_b)$.


%% file: Sections/corrections.tex
\newcommand{\CKKs}{C^{\rm KK}_s}
\newcommand{\CKKb}{C^{\rm KK}_b}
\newcommand{\Clogs}{C^{\rm log}_s}
\newcommand{\Cxi}{C^{\xi}_1}
\newcommand{\Cxii}{C^{\xi}_2}
\newcommand{\Cflux}{C^{\rm flux}}
\newcommand{\Ccon}{C^{\rm con}}
\newcommand{\CF}{C^F}

\newcommand{\overall}{\left(\frac{g_s^4}{8\pi||\Omega||^2}\right)^{-1}}

\newpage
\section{Dangerous corrections?}
\label{sec:dangerous-corrections}

Although there seem to exist explicit constructions of dS vacua with an $\antiD$-brane uplift in the context of LVS, both in the weakly-warped regime \cite{Bento:2021nbb}, as we just reviewed,  and in the strongly-warped one \cite{LVSdS:2010.15903}, these implicitly rely on the assumption that all putative subleading corrections are indeed subleading compared to the contributions that have already been included in the 4d low energy effective field theory.  The question of whether this assumption is justified was analysed in  \cite{Junghans:2022exo} for the strongly-warped regime studied in \cite{LVSdS:2010.15903}, by taking into account various types of corrections to the LVS potential.  The conclusion for the strongly-warped regime was that some corrections do not have any parametric suppression, and moreover even those that do can never be \emph{simultaneously} parametrically suppressed.

It is worth fleshing out the main point in the argument of \cite{Junghans:2022exo}. If a correction to the LVS potential can arise from some string theory effect (e.g.~one-loop or curvature corrections), one of two things must happen: either this correction is parametrically suppressed for all quantities in the solution of interest (moduli vevs, masses, vacuum energy), and can therefore be neglected in the limit of weak coupling and/or large volume within the EFT, or else one \emph{must} compute this correction explicitly and include it in the analysis, and potentially worry about next-to-next-to-leading order corrections.  While the result of \cite{Junghans:2022exo} does not show that finding a stable strongly-warped dS solution in LVS is impossible, it means that one must take at least some of these corrections into account, compute their coefficients explicitly, and keep them in the EFT unless their numerical coefficients can be made small.  This constitutes a challenge because, not only are these coefficients typically hard to compute explicitly, but also the known solutions in the LVS may be absent or at least corrected in significant ways by these contributions.

It therefore becomes relevant to ask whether this is also the case in the weakly-warped de Sitter vacua just discussed \cite{Bento:2021nbb}.  In order to address this question, we will briefly outline the known subleading contributions to the  4d low energy effective theory (for more details see \cite{Junghans:2022exo} and references therein) and compute how they correct the off-shell scalar potential in the weakly-warped regime.  We will then investigate the consequences of these corrections for the moduli vevs and vacuum energy, and work out how they can be suppressed in order to achieve self-consistent weakly-warped LVS dS vacua.

\subsection{Corrections to the off-shell potential}

Let us group the known subleading contributions to the 4d EFT according to their origin and work out how they correct the off-shell scalar potential in the weakly-warped regime.

\subsection*{Curvature/loop corrections to the K\"ahler potential}

\noindent String-loop corrections at order $\alpha'^2$ from the exchange of KK modes between D7/D3 branes or O7/O3 planes \cite{Berg:2005ja,Berg:2014ama,Haack:2015pbv,Haack:2018ufg,Garcia-Etxebarria:2012bio, Gao:2022uop} enter the K\"ahler potential \eqref{eq:LVS_Kahler_w} as \cite{Cicoli:2007xp,Berg:2007wt,vonGersdorff:2005bf,Berg:2005yu}
\begin{align}
    \delta K = \CKKs\cdot\frac{g_s^2 \sqrt{\tau_s}}{\V}
    + \CKKb\cdot\frac{g_s^2 \sqrt{\tau_b}}{\V} \,,
\end{align}
and thus the scalar potential is corrected as (cf. first line of eq.~\eqref{E:largebetaV})
\begin{align}
    \overall\delta V =&~ \CKKs\cdot \frac{g_s^2}{6\kappa_s\tau_s}\cdot\frac{8a^2A^2 \sqrt{\tau_s} e^{-2\frac{a\tau_s}{g_s}}}{3g_s^2\kappa_s\V}
    +\CKKs\cdot \frac{g_s^2}{3\kappa_s\tau_s} 
    \cdot\frac{4 a A W_0 \tau_s e^{-\frac{a\tau_s}{g_s}}}{g_s\V^2} \nonumber \\
    &+\left\{(\CKKs)^2\cdot \frac{2g_s^4}{9\kappa_s\sqrt{\tau_s}}
    + \CKKb\cdot \frac{8c' g_s^2 (g_s M)^2\zeta^{2/3}}{\V^{1/3}}
    \right\}
    \cdot\frac{3W_0^2}{4\V^3} \,.
\end{align}
Note that in principle the coefficients $\{\CKKs,\CKKb\}$ depend on the complex structure moduli; however, as this dependence is unkown, we will simply assume that $\CKKs,\CKKb\sim \Op{1}$ and neglect the derivative contributions.
For easy identification, we have written each term isolating clearly each suppression factor with respect to the corresponding uncorrected term in the LVS potential. For example, all the terms above are suppressed by a factor of $g_s^2$ compared to their LVS counterparts (see eq.~\eqref{E:largebetaV}). The $\CKKb$ term is also suppressed by the ratio $\zeta^{2/3}/\V^{1/3}$. 

On the other hand, threshold corrections to gauge couplings lead to the one-loop field redefinition of the K\"ahler modulus
$\tau_s^{\rm new} = \tau_s^{\rm old} + C^{\rm log}_s \log\V$, which implies a correction to the K\"ahler potential once the physical volume is expressed in terms of the corrected modulus, $\V = \tau_b^{3/2} - \kappa_s(\tau_s - \Clogs\log\V)^{3/2}$ \cite{Conlon:2010ji}. It contributes to the scalar potential as\footnote{This field redefinition leads to a non-linear relation between the volume $\V$ and the K\"ahler moduli ($\tau_b,\tau_s$). Treating this redefinition as a small correction, one can solve it perturbatively for $\V = \tau_b^{3/2} - \kappa_s\tau_s^{3/2} + \Op{\Clogs}$ and use this leading order solution in the correction. Hence  at leading order  the volume becomes $\V\approx \tau_b^{3/2} - \kappa_s\Big(\tau_s - \Clogs\log(\tau_b^{3/2} - \kappa_s^{3/2})\Big)^{3/2}$.} 
\begin{align}\label{eq:Clogs}
    \overall\delta V =&~ \Clogs\cdot \frac{g_s\log\V}{2\tau_s}
    \cdot\frac{8a^2A^2 \sqrt{\tau_s} e^{-2\frac{a\tau_s}{g_s}}}{3g_s^2\kappa_s\V}
    + \Clogs\cdot \frac{g_s\log\V}{\tau_s}
    \cdot\frac{4 a A W_0 \tau_s e^{-\frac{a\tau_s}{g_s}}}{g_s\V^2} \nonumber \\
    &- \Clogs\cdot 6 g_s\kappa_s\sqrt{\tau_s}
    \cdot\frac{3W_0^2}{4\V^3} \,.
\end{align}
Once again, notice the relative factors with respect to the terms in the LVS potential \eqref{E:largebetaV}. Recalling that the LVS solution is $\log\V\sim \frac{a\tau_s}{g_s}$, there seems to be no parametric suppression in the correction terms in the first line.

At order $\alpha'^3$ there are corrections from different sources \cite{Berg:2014ama,Haack:2015pbv,Haack:2018ufg,Minasian:2015bxa,Antoniadis:2019rkh} (such as the backreaction of 7-branes \cite{Minasian:2015bxa} and the exchange of KK modes with D7 branes/O7 planes \cite{Antoniadis:2019rkh}), which contribute at leading order as a redefinition of $\xi\to\xi - \Delta\xi + \Cxi g_s\log\V + \Cxii g_s$ in $K$ \eqref{eq:LVS_Kahler_w}, correcting the scalar potential as
\begin{align}\label{eq:Cxi}
    \overall\delta V =&~ (-\Delta\xi + \Cxi g_s\log\V + \Cxii g_s)\cdot\frac{3W_0^2}{4\V^3} \,.
\end{align}
Notice again that there seems to be no suppression for the $\Delta\xi$ and the $\Cxi$ corrections. 
However, $\Delta\xi$ can be computed explicitly in terms of the Poincar\'e dual $D_{\rm O7}$ of the divisor wrapped by O7 planes \cite{Minasian:2015bxa},
\begin{equation}
    \Delta\xi = \frac{\zeta(3)}{(2\pi)^3}\int_X D_{\rm O7}^3 \,,
\end{equation} 
and it amounts simply to a shift in the numerical parameter $\xi$.

\subsection*{Curvature corrections to the gauge-kinetic function}
The non-perturbative superpotential which is a key ingredient in the stabilisation of the K\"ahler moduli arises from gaugino condensation on a stack of D7-branes \cite{Jockers:2004yj,Jockers:2005zy,Haack:2006cy}. Therefore curvature corrections to the D7-brane action \cite{Bachas:1999um,Bershadsky:1995qy,Green:1996dd,Cheung:1997az,Minasian:1997mm}, which correct the gauge-kinetic function of the D7-branes will affect this non-perturbative contribution. The gauge-kinetic function becomes $f_s = T_s - \frac{\chi_s}{24}\tau$, where $\chi_s$ is the Euler number of the 4-cycle wrapped by the D7-branes whose size is controlled by $\tau_s$, so that the generated superpotential becomes
\begin{align}\label{eq:Wnp_correction}
    W_{np} = A\cdot e^{-\frac{a}{g_s}\tau_s + a\frac{\chi_s}{24g_s}} \,,
\end{align}
which is equivalent to a redefinition of $A\to A\cdot e^{\frac{a\chi_s}{24g_s}}$. It is worth recalling that our weakly-warped solution required somewhat large values of the parameter $A$ (\refeq{eq:AW0_bound}) --- one might therefore expect that this correction in particular will benefit, rather than obstruct, these weakly-warped solutions. 

\subsection*{Higher F-terms}
Further corrections arise from higher F-terms associated with 4 superspace derivatives in the 4d EFT \cite{deAlwis:2012vp,Cicoli:2013swa,Ciupke:2015msa}. These could come for example from integrating out KK modes \cite{Cicoli:2013swa} or from 10d 8-derivative terms with powers of $G_3$ \cite{Ciupke:2015msa}, which would lead to such a contribution in the potential. The scalar potential gets corrected as
\begin{align}\label{eq:CF}
    \overall\delta V = \CF\cdot \frac{8g_s^2 W_0^2}{3\V^{2/3}}
    \cdot \frac{3W_0^2}{4\V^3} \,.
\end{align}
This correction is suppressed relative to the LVS potential by the factor 
\begin{align}
    \frac{m_{3/2}^2}{m_{KK}^2} \sim \frac{g_s^2 W_0^2}{\V^{2/3}} \,,
\end{align}
which one would generically want to be small in order to keep the gravitino in the EFT valid at energies $E\ll m_{KK}$ (and therefore preserve $\mathcal{N} = 1$ supersymmetry).  

\vskip 1em

So far only one of the corrections is warping-related, the correction linear in $\CKKb$ which depends on the warping due to its dependence on $\zeta$. It does indeed arise from the warping correction to the deformation modulus metric, which mixes this modulus with the volume $\V$ --- this was precisely the crucial term in the discussion of our weakly-warped solutions. One   would not expect the effect of the other corrections thus-far considered to be dependent on which warping regime we consider, although, as we will see below,  having weak-warping rather than strong-warping can still make a difference due to the different region of interest in parameter space.  
The two remaining corrections to be considered, $\{\Cflux, \Ccon\}$, are intrinsically warping-related, being associated with the tip of the warped conifold, and one would therefore expect them to play a more important role in the distinction between the two regimes.

\vskip 1em
\subsection*{Curvature/warping corrections from conifold-flux backreaction}
The charges responsible for the warping (such as branes, O-planes and fluxes) can have large contributions to the 10d curvature even at large volumes and result in large curvature corrections \cite{Gao:2020xqh,Carta:2019rhx}. For the KS throat, the charge responsible is the flux contribution $KM$, which in the strongly-warped regime is required to be large in order that the warping is sufficient to avoid runaway in the volume modulus after uplifting. This leads to corrections proportional to the ratio $\frac{KM}{\V^{2/3}}$, e.g. from dimensionally reducing the 10d $R^4$ terms of type IIB, which appear in the scalar potential as \cite{Junghans:2022exo}
\begin{align}\label{eq:Cflux}
    \overall\delta V = \Cflux\cdot \frac{5 g_s K M}{\V^{2/3}}
    \cdot\frac{3W_0^2}{4\V^3} \,.
\end{align}
Since in the weakly-warped regime, one does not require very large charges $KM$, this correction is expected to be less dangerous than in the strongly-warped case.\footnote{The form of the correction used in \cite{Gao:2022fdi} for the LVS parametric tadpole constraint differs by a factor of $\xi$, which is just $\bigo{1}$ in our example. The discrepancy follows from the assumption in \cite{Gao:2022fdi} that the warp factor is a slowly varying function, in which case one should expect this warping correction to scale with the Euler number $\chi(X_6)$. See, however, \cite{Junghans:2022kxg} for further discussion.}

\subsection*{Curvature corrections in the conifold region}
Since the radius of the $S^3$ at the tip of the conifold is $R^2_{S^3}\sim (2\pi)^2\V^{1/3}\zeta^{2/3} \alpha'$, one expects $\alpha'$ curvature corrections to be suppressed by $((2\pi)^2\V^{1/3}\zeta^{2/3})^{-1}$, in contrast with the strongly-warped suppression by $(g_sM)^{-1}$. E.g. the $R^2$ corrections to the DBI action of the $\antiD$-brane in the warped deformed conifold background correct the brane tension\footnote{These $(\alpha')^2$ curvature corrections to the brane action were studied in \cite{Hebecker:2022zme,Schreyer:2022len} and found to lower the tension of the brane, providing an alternative for warped throats.} {with a term suppressed by $((2\pi)^2\V^{1/3}\zeta^{2/3})^{-2}$}.  In practice, this corresponds to the redefinition  
\begin{align}
    \cD \to \cD \Big(1 + \frac{\Ccon}{(g_s M)^2}\frac{1}{\beta^3}\Big) \,,
\end{align}
which is suppressed by $1/\beta^3$ with respect to the leading order potential in the limit $\beta\gg 1$.

Let us now look at the solutions for the corrected weakly-warped LVS potential, as well as at the vacuum energy associated with these vevs, and examine the suppression of each correction on these quantities.

\subsection{Corrections to the solutions}

We now compute how the solutions are modified due to the various corrections outlined above. 

\subsubsection{Conifold modulus stabilisation}

The only corrections involving the conifold modulus, and therefore affecting its stabilisation, are those  associated with $\{\CKKb,\Ccon\}$. The $\Ccon$ correction is suppressed by $1/\beta^3$ in the limit $\beta\gg 1$ and will therefore not be dangerous for the solution in the weakly-warped regime. 
The corrections {$\CKKb$} can be taken into account by solving perturbatively in a similar way to how the $1/\beta$ corrections were taken into account.  We find that the vev of $\zeta$ gets corrected as (see \eqref{eq:solz_largebeta})
\begin{align}\label{eq:solz_corrected}
    \zeta \approx 
    \zGKP
    \Bigg(
        1 &- \cD\frac{4 K}{3\pi^2 c'' M ~ \zGKP^{4/3}\V^{2/3}} 
        + \Ccon\cdot \Op{\frac{1}{\beta^3}} \nonumber \\
        &- C^{\rm KK}_b\cdot \frac{2c'(2\pi)^2}{\pi||\Omega||^2}\frac{2\pi K}{g_s M}\,\zGKP^{2/3}\, \frac{g_s^4W_0^2}{\V^{4/3}}
    \Bigg) \,,
\end{align}
that is, the $\CKKb$ corrections are heavily suppressed by the factor 
$\frac{\zGKP^{2/3}}{\V^{2/3}}\cdot\frac{g_s^4W_0^2}{\V^{2/3}}$. 

\subsubsection{K\"ahler moduli}

Since the formal solution for $\V$ follows directly from $\partial_{\tau_s} V = 0$, one can also infer that only a few of the corrections will appear --- in particular, it will receive explicit corrections from $\{\CKKs,\Clogs,\chi_s\}$ only. Of course, due to its dependence on $\tau_s$, which is sensitive to all corrections considered, $\V$ is also implicitly sensitive to all corrections. This is also true for $\zeta$, since the subleading terms in its vev (\refeq{eq:solz_corrected}) depend on $\V$.

Solving perturbatively to leading order in all corrections, we find (cf. \eqref{eq:solV}, \eqref{eq:soltaus_largebeta})
\begin{align} \label{eq:solv_corrected}
    \V &\approx \Bigg(
        \frac{3(a\tau_s - g_s)}{4 a \tau_s - g_s} 
        + \CKKs\cdot \frac{g_s^2}{8\kappa_s \tau_s}
        - \Clogs\cdot \frac{9a}{8}    
    \Bigg)
        \frac{g_s W_0 \kappa_s \sqrt{\tau_s}}{a A}
        \cdot e^{\frac{a}{g_s}\tau_s - a\frac{\chi_s}{24g_s}}\,, \\
    \label{eq:soltaus_corrected}
    \tau_s &\approx \frac{\hat{\xi}^{2/3}}{(2\kappa_s)^{2/3}} 
    + \frac{(1+2\alpha) g_s}{3a} 
    + \Op{\frac{1}{\beta}} + \Op{g_s^2} \nonumber \\ 
    &-\CKKs \cdot\frac{g_s^2}{3\kappa_s}
    +\Clogs \cdot\frac{5a}{3}\Big(\frac{\hat{\xi}}{2\kappa_s}\Big)^{2/3}
    +\Cxi \cdot\frac{a}{3\kappa_s}\Big(\frac{\hat{\xi}}{2\kappa_s}\Big)^{1/3}
    +\Cxii \cdot\frac{g_s}{3\kappa_s}\Big(\frac{\hat{\xi}}{2\kappa_s}\Big)^{-1/3} \nonumber \\
    &
    +\Cflux \cdot\frac{55}{27\kappa_s}\frac{g_s KM}{\V^{2/3}}\Big(\frac{\hat{\xi}}{2\kappa_s }\Big)^{-1/3} 
    +\CF \cdot\frac{88}{81\kappa_s }\frac{g_s^2 W_0^2}{\V^{2/3}}\Big(\frac{\hat{\xi}}{2\kappa_s}\Big)^{-1/3} \nonumber \\
    %
    %
    &+\CKKb\cdot 
    \frac{80c'(g_sM)^2}{27\kappa_s}\cdot\frac{g_s^2 \zeta^{2/3}}{\V^{1/3}}\Big(\frac{\hat{\xi}}{2\kappa_s}\Big)^{-1/3}\,,
\end{align}
with $\alpha$ given by (\ref{eq:alpha}) and $\hat{\xi}\equiv \xi - \Delta\xi$.
The solution for $\V$ (\refeq{eq:solv_corrected}) matches\footnote{Apart from a factor of $3$ in the $\Clogs$ term.} the one found in \cite{Junghans:2022exo} up to factors of $g_s$ due to the use of different conventions (see below).
This agrees with our expectation that the explicit form of the $\{\CKKs,\Clogs,\chi_s\}$ corrections is independent of the warping regime.
One can immediately see that, although the $\CKKs$ correction is suppressed by $g_s^2$, neither the $\Clogs$ nor the $\chi_s$ corrections are suppressed --- following the logic presented above, one must therefore include these corrections in the EFT analysis. Although the $\chi_s$ correction is not only not suppressed but actually enhanced by a factor of $1/g_s$, the fact that we know its explicit form means that it can be taken into account when looking for a solution and, as we saw in (\refeq{eq:Wnp_correction}), it can be seen as a rescaling of the parameter $A$ in the non-perturbative contribution to the superpotential. In fact, given the need for exponentially large $AW_0$ \eqref{eq:AW0_bound}, this seems to help rather than hurt weakly-warped solutions.\footnote{Indeed, for the CY$_3$ of \cite{LVSdS:2010.15903}, we have $\chi_s=3$ and hence the correction represents a rescaling of $A\to A e^{\frac{a\chi_s}{24 g_s}}$, which would na\"ively allow for the lower value $A\sim 403$.} The $\Clogs$ correction, on the other hand, is not generically known explicitly and therefore one cannot follow the same procedure --- this coefficient would have to be computed for the compact geometry of interest in order to be included in the EFT analysis. 

The different $g_s$ factors in the solution with respect to \cite{Junghans:2022exo} arise from our choice of frame conventions, with $\tau_b$ and $\tau_s$ differing from the ones in \cite{Junghans:2022exo} by a factor of $g_s$, i.e. $\tau_i^{\rm (ours)} = g_s\tau_i^{\text{\cite{Junghans:2022exo}}}$ and $\V^{\rm (ours)} = g_s^{3/2}\V^{\text{\cite{Junghans:2022exo}}}$. 
Recall that in these conventions, the Einstein-frame volumes and the string-frame volumes are the same at the vev, so that deciding e.g. whether $\alpha'$-corrections are under control for a given solution can be done directly in terms of the Einstein-frame volumes.\footnote{Note that checking whether the $\alpha'$-corrections are under control, i.e.~whether a given solution is consistent with the supergravity description being used, should be done using string-frame volumes, which are parametrically (in $g_s$) smaller than the Einstein-frame volumes when the convention $\tau_i^{(S)} = g_s\tau_i^{(E)}$ is used for the 10d change of frames, rather than the one we are using in this work for which these volumes are the same. Therefore, one might naively think that the volumes are large ``enough'' in Einstein-frame, while the string-frame volumes are actually inconsistent with the $\alpha'$-expansion.}
It also makes explicit the $1/g_s$ factor in the non-perturbative superpotential, rather than it appearing in the solution for $\tau_s$ --- both pictures are ultimately consistent since this factor of $1/g_s$ will appear in the potential one way or another. Because of this, however, our solution for $\tau_s$ (\ref{eq:soltaus_corrected}) differs from the one in \cite{Junghans:2022exo} by an overall factor of $g_s$, as can be most easily seen by looking at the terms that behave similarly in both warping regimes. 

The first line of (\ref{eq:soltaus_corrected}) shows the expected LVS solution for $\tau_s$ which already includes an $\Op{g_s}$ correction (cf. \eqref{eq:soltaus_largebeta}), with the $\antiD$-brane backreaction on $\tau_s$ encoded by the uplift parameter $\alpha$. 
The corrections in the second and third lines are suppressed in the same way as in the strongly-warped regime\footnote{The $\Clogs$ factor again differs by a factor of $5/3$.}, as expected from the fact that they do not involve the interplay between $\zeta$ and $\V$. Whereas the $\{\CKKs,\Cflux,\CF\}$ terms are suppressed by powers of $g_s$ and/or $\V$, $\{\Clogs,\Cxi\}$ have no suppression whatsoever, and $\{\Cxii\}$ is only suppressed by one power of $g_s$; the latter is still dangerous because of the way the vev of the volume $\V$ depends on $\tau_s$, i.e. 
\begin{align}
    \V \propto e^{\frac{a}{g_s}\tau_s} \approx e^{\frac{a}{g_s}\tau_s^{(0)}}\cdot\exp
    \bigg(1 + \Cxii \frac{a}{3\kappa_s}\Big(\frac{\hat{\xi}}{2\kappa_s}\Big)^{-1} + ... \bigg) \,.
\end{align}
It also follows that the $\{\Clogs,\Cxi\}$ terms will both appear with a $1/g_s$ power relative to the leading solution and therefore grow at small couplings.

Finally, the $\Ccon$ correction is suppressed in large $\beta$ and hence not dangerous in the weakly-warped regime.

\subsubsection{Vacuum Energy}

With the vevs of the moduli in (\refeq{eq:solz_corrected}--\refeq{eq:soltaus_corrected}), we can compute the vacuum energy given by the potential $V$ at the minimum, and look at the effect of each correction. We can write it as (see \eqref{eq:Vmin})
\begin{align}\label{eq:Vmin-overall}
    V_{min} = \frac{g_s^4}{8\pi||\Omega||^2} \cdot 
    \frac{3g_s \kappa_s W_0^2 \sqrt{\tau_s}}{4a\V^3} \rho\,,
\end{align}
with $\rho$ defined as
\begin{align}\label{eq:Vmin-corrected}
    \rho =& ~\alpha - 1 + \Op{g_s} + \Op{1/\beta} \\
    &-\CKKs\cdot \frac{g_s^2}{6 \kappa_s} \Big(\frac{\hat{\xi}}{2\kappa_s}\Big)^{-2/3}
    +\Clogs\cdot \frac{a}{6}(11+12\log\nu) 
    +\Cxi\cdot \frac{a \log\nu}{\kappa_s} \Big(\frac{\hat{\xi}}{2\kappa_s}\Big)^{-1/3}
    -\Cxii\cdot \frac{g_s}{3\hat{\xi}}
    \nonumber \\
    &
    -\Cflux\cdot \frac{10a}{9 \kappa_s }\frac{KM}{\V^{2/3}} \Big(\frac{\hat{\xi}}{2\kappa_s}\Big)^{-1/3} 
    -\CF\cdot \frac{16a}{27\kappa_s}\frac{g_s W_0^2}{\V^{2/3}}\Big(\frac{\hat{\xi}}{2\kappa_s}\Big)^{-1/3}
    \nonumber \\ 
    &
    %
    -\CKKb\cdot \frac{8c' a (g_sM)^2}{9\kappa_s}\frac{g_s \zeta^{2/3}}{\V^{1/3}}
    \Big(\frac{\hat{\xi}}{2\kappa_s}\Big)^{-1/3}
     \nonumber \,.
\end{align}
where 
\begin{equation}
    \nu = \frac{3g_s \kappa_s W_0 }{8 a A}\left(\frac{\hat{\xi}}{2\kappa_s}\right)^{1/3} \,. 
\end{equation}

The first line reproduces the leading order result \eqref{eq:Vmin} and requires $\alpha > 1$ provided all corrections can be safely neglected.
The correction terms in the second and third lines are suppressed by the same factors as in the strongly-warped case. In particular, the terms $\{\Clogs,\Cxi\}$ appear unsuppressed and are thus the most dangerous. On the other hand the $\{\CKKb\}$ correction is suppressed by factors of $\zeta,g_s\ll 1$ and $\V\gg 1$, suggesting that it will not be among the most dangerous for the solution.

Note that the non-perturbative no-scale behaviour \cite{Junghans:2022exo} is manifesting itself in the way the factors $\{\Cflux,\CF\}$ are suppressed relative to the leading LVS solution in $\rho$ compared to their suppression in the off-shell potential, \eqref{eq:CF} and \eqref{eq:Cflux} respectively, having acquired a $1/g_s$ enhancement in the vacuum energy (see discussion below). It is also worth pointing out that the unsuppressed corrections $\{\Clogs,\Cxi,\Cxii\}$ are independent of the presence of the conifold modulus and the brane --- these are potentially dangerous corrections even for pure LVS.

One might wonder whether it  is possible to choose a specific model (i.e.~a specific compact space geometry) such that the unsuppressed corrections to the vevs and vacuum energy $\{\Clogs,\Cxi,\Cxii\}$ are not present,  the remaining corrections are all consistently suppressed,  and the solution is parametrically under control.  Note that the $\CKKs$ correction is automatically suppressed as long as the string loop expansion $(g_s\ll 1)$ is under control, which we must require for consistency of our analysis. Suppression of the $\CKKb$ correction,
	\begin{align}
		(g_sM)^2 \frac{g_s\zeta^{2/3}}{\V^{1/3}}
		\ll 1 \,,
	\end{align}
is also easily satisfied in the weakly-warped region of parameter space. 

In order to suppress the correction $\{\Cflux\}$, we require that
\begin{align}
    \frac{KM}{\V^{2/3}} \ll 1 \,. \label{eq:Cfluxsmall}
\end{align}
This condition is also easily satisfied, but it does have interesting implications.  Firstly, recall that a solution in the weakly-warped regime must also satisfy \eqref{eq:W0V-bound}, which gives, parametrically,\footnote{Note that although this condition from \eqref{eq:W0V-bound} may seem to imply that we need $W_0 \gg 1$, and not just $A W_0 \gg 1$, when replacing the volume modulus vev \eqref{eq:solV}, $\mathcal{V} \sim W_0 g_s/A$, we recover \eqref{eq:AW0_bound} and the large $A W_0$ bound.}
\begin{align}
    \frac{\V}{g_s^2  W_0^2} < \bigo{1} \,. \label{eq:VgsW0}
\end{align}
Using Eq. (\ref{eq:VgsW0}), together with the bound \eqref{eq:DDbound},  on the condition that $\Cflux$ corrections be small \eqref{eq:Cfluxsmall} implies\footnote{We thank Arthur Hebecker for pointing this out to us.}
\begin{equation}
    1 \gg \frac{KM}{\V^{2/3}} > \frac{KM}{(g_s^{2}W_0^{2})^{2/3}} > \frac{KM}{g_s^{2/3}N_\text{flux}^{2/3}} \,.
\end{equation}
Thus we must have large bulk flux contributions to the D3-tadpole, $\Nflux \gg MK$, which is consistent with the need for large $W_0$.

Finally, parametric suppression of the final correction $\{C^F\}$ requires
\begin{align}
	\frac{g_s W_0^2}{\V^{2/3}} \ll 1 \,. \label{eq:CFsmall}
\end{align}
This is in tension with the weakly-warped regime; together \eqref{eq:CFsmall} and \eqref{eq:VgsW0} imply
\begin{align}
    \V^{1/3} \ll g_s \,,
\end{align} 
which is never consistent. One can check explicitly using the parameter set of Table \ref{tb:parlargebeta} that  taking into account all numerical coefficients $\{a,\kappa_s,\hat{\xi},||\Omega||^2\}$ does not change the conclusion. Therefore the $\CF$ correction is unsuppressed unless $\CF\ll 1$.

It is worth noting that already in the strongly-warped case, the $\CF$ correction is dangerous for small $g_s$, which may not be obvious from the suppression factor $\propto g_s W_0^2/\V^{2/3}$. One would na\"ively expect that since $\V\propto g_s\cdot e^{1/g_s}$, smaller values of $g_s$ would suppress this correction the strongest. However, it is important to note that one is not free to choose $W_0$ independently --- in particular, in the strongly-warped case \cite{Junghans:2022exo} (adapted to our conventions)
\begin{equation}
    W_0^2 \propto \frac{\zeta^{2/3}\V^{5/3}}{\tilde{\alpha}~g_s^3} 
    \implies 
    W_0^{1/3} \propto \frac{\zeta^{2/3}}{\tilde{\alpha}~g_s^{4/3}}\cdot e^{\frac{5}{3}\frac{a}{g_s}\tau_s}
    \,,
\end{equation} 
with $\tilde{\alpha}\sim \Op{1}$. Plugging this into the $\CF$ suppression factor, 
\begin{equation}
    \frac{g_s W_0^2}{\V^{2/3}} \sim \frac{\zeta^{8/3}}{\tilde{\alpha}^4 g_s^{5}} e^{6\cdot \frac{a}{g_s}\tau_s} \,,
\end{equation}
shows that indeed it grows for small $g_s$ contrary to our initial expectation. At the same time, this growth can be compensated by exponentially small $\zeta$, which can be achieved through large $MK$ consistently with \cite{Gao:2022fdi,Junghans:2022kxg}. 

Just as in the strongly-warped regime, one concludes that several corrections could be dangerous in the context of weakly-warped LVS de Sitter vacua. Parametric suppression of the $\Cflux$ correction would need large bulk flux contributions to the D3-tadpole; otherwise one would need the numerical coefficient $\Cflux$ to be small. One would at least need the corrections $\{\Clogs,\Cxi,\Cxii,\CF\}$ to either vanish or be suppressed through the explicit form of these numerical coefficients.  This would require explicit computation of these coefficients in order to choose a construction for which they can be small, though they are not currently under control for general Calabi-Yau geometries and setups.  Alternatively, if they are $\mathcal{O}(1)$,  then the corresponding corrections would need to be included in the 4d low energy effective field theory and new searches for de Sitter solutions would need to be performed.

\subsection{Comment on the NPNS}

In \cite{Junghans:2022exo}, a property of the LVS potential called the non-perturbative no-scale (NPNS) structure was emphasised due to its effect on the balance between leading and subleading (correction) terms in a given solution. 
The root of the NPNS is the fact that the minimum in LVS --- which is to be considered the leading contribution to which subleading corrections are added --- itself appears at subleading order in $g_s$. To see this, consider the change of variables $\V = \nu\cdot g_s \cdot e^{t/g_s}$ and $\tau_s=t/a$, for which the LVS potential takes the form 
\begin{equation}\label{eq:NPNS-Voffshell}
    V = g_s\cdot e^{-\frac{3}{g_s}t} f(t,\nu) \,, 
\end{equation}
where $f(t,\nu)$ does not depend on $g_s$.\footnote{
    Explicitly, we have 
    \begin{equation}
        f(t,\nu) = 
        \frac{8a^{3/2}A^2\sqrt{t}}{3\kappa_s \nu}
        - \frac{4 A W_0 t }{\nu^2}
        +\frac{3W_0^2}{4\nu^3}\xi \,. 
    \end{equation} 
    One might wonder whether $W_0$ and $A$ in particular can be important for the argument we outline below. Note however that rescaling $\nu \to \frac{W_0}{A} \cdot\nu$, the function becomes  
    \begin{equation}
        f(t,\nu) = \frac{A^3}{W_0}\left\{
        \frac{8a^{3/2}\sqrt{t}}{3\kappa_s \nu}
        - \frac{4 t }{\nu^2}
        +\frac{3\xi}{4\nu^3} \right\} \,. 
    \end{equation} 
    Since the ratio $A^3/W_0$ only appears as an overall factor, these parameters will not affect the argument for the NPNS.
} A minimum of this potential therefore requires 
\begin{align}\label{eq:NPNS-pdt}
    \partial_t V &= -\frac{3}{g_s}\cdot g_s\cdot e^{-\frac{3}{g_s}t}f(t,\nu) + g_s\cdot e^{-\frac{3}{g_s}t}(\partial_t f)  \\ 
    &= -\frac{3}{g_s}V + g_s\cdot e^{-\frac{3}{g_s}t}(\partial_t f)
    \overset{!}{=} 0 \,,
\end{align}
which implies that the scalar potential at the minimum is suppressed by a factor of $g_s$ compared to the off-shell potential \eqref{eq:NPNS-Voffshell}, 
\begin{equation}\label{eq:NPNS-Vmin}
    V_{min} = \frac{g_s}{3}\cdot g_s\cdot e^{-\frac{3}{g_s}t} (\partial_t f) \,.
\end{equation}
Another way of thinking about this result, which is useful once we add corrections, is that solving equation \eqref{eq:NPNS-pdt} perturbatively in $g_s$, i.e. $t = t_0 + t_1 g_s + \Op{g_s^2}$, the leading order solution $t=t_0$ would result in a vanishing potential (which would in turn leave $\nu$ as a flat direction). 

Now consider adding a correction suppressed by $\epsilon\cdot e^{-\frac{\lambda}{g_s}t}$,
\begin{equation}\label{eq:NPNS-Voffshell-corrected}
    V = g_s\cdot e^{-\frac{3}{g_s}t} f(t,\nu)
    + g_s\cdot\epsilon\cdot e^{-\frac{\lambda}{g_s}t} g(t,\nu) \,,
\end{equation}
where $g(t,\nu)$ does not contain $g_s$ or any other small parameters.  Minimising the potential now requires 
\begin{align}\label{eq:NPNS-pdt-corrected}
    \partial_t V =& -\frac{3}{g_s}\cdot g_s\cdot e^{-\frac{3}{g_s}t}f(t,\nu) + g_s\cdot e^{-\frac{3}{g_s}t}(\partial_t f) \\
    &-\frac{\lambda}{g_s}\cdot g_s\cdot\epsilon\cdot e^{-\frac{\lambda}{g_s}t} g(t,\nu) 
    + g_s\cdot\epsilon\cdot e^{-\frac{\lambda}{g_s}t}(\partial_t g) \nonumber \\
    =& -\frac{3}{g_s}\cdot\left\{
        g_s\cdot e^{-\frac{3}{g_s}t}f(t,\nu) 
        + g_s\cdot\epsilon\cdot e^{-\frac{\lambda}{g_s}t} g(t,\nu)
    \right\} \\
    &+g_s\cdot e^{-\frac{3}{g_s}t}(\partial_t f)
    + g_s\cdot\epsilon\cdot  e^{-\frac{\lambda}{g_s}t}(\partial_t g) 
    +(3-\lambda)\cdot\epsilon\cdot e^{-\frac{\lambda}{g_s}t} \cdot g(t,\nu) \nonumber \\ 
    =& -\frac{3}{g_s}V 
    +g_s\cdot e^{-\frac{3}{g_s}t}(\partial_t f)
    + g_s\cdot\epsilon\cdot  e^{-\frac{\lambda}{g_s}t}(\partial_t g) \nonumber \\
    &+(3-\lambda)\cdot\epsilon\cdot e^{-\frac{\lambda}{g_s}t} \cdot g(t,\nu)
    \overset{!}{=} 0 \,,
\end{align}
which now implies for the vacuum energy
\begin{align}\label{eq:NPNS-Vmin-corrected}
    V_{min} =& 
        \frac{g_s^2}{3}\cdot e^{-\frac{3}{g_s}t}(\partial_t f)
        +\frac{g_s^2}{3}\cdot\epsilon\cdot  e^{-\frac{\lambda}{g_s}t}(\partial_t g) 
        +\frac{3-\lambda}{3}\cdot g_s\cdot\epsilon\cdot e^{-\frac{\lambda}{g_s}t} \cdot g(t,\nu)
    \,.
\end{align}
Note that, unless $\lambda=3$ (i.e.~the correction satisfies the NPNS structure), the last term is the leading contribution from the corrections and \emph{it is not suppressed by the extra factor of} $g_s$ that appears in the first two terms --- the extra $g_s$ suppression characteristic of LVS \eqref{eq:NPNS-Vmin}. Therefore, to safely neglect this contribution from the correction one should have\footnote{Note that the suppression factor must include the exponential $e^{-\frac{\lambda-3}{g_s}t}$, which corresponds to the volume suppression (whose power is given by $\lambda-3$). In practice, $\epsilon$ counts all factors apart from powers of the volume.}
\begin{align}\label{eq:NPNS-condition}
    g_s\cdot\epsilon\cdot e^{-\frac{\lambda}{g_s}t}\ll g_s^2\cdot e^{-\frac{3}{g_s}t} 
    \quad\Rightarrow\quad
    \frac{\epsilon}{g_s}\cdot e^{-\frac{(\lambda-3)}{g_s}t} \ll 1  \,, 
\end{align}
rather than the naively expected $\epsilon\cdot e^{-\frac{(\lambda-3)}{g_s}t}\ll 1$.

To compare with \eqref{eq:Vmin-corrected} one should look at $\rho$, by taking out the suppression factors $g_s^2\cdot e^{-\frac{3}{g_s}t}$ as in \eqref{eq:Vmin},
\begin{equation}
    \rho\propto 
    \frac{1}{3}(\partial_t f)
    +\frac{1}{3}\cdot\epsilon\cdot  e^{-\frac{(\lambda-3)}{g_s}t}(\partial_t g) 
    +\frac{(3-\lambda)}{3}\cdot \frac{\epsilon}{g_s}\cdot e^{-\frac{(\lambda-3)}{g_s}t} \cdot g(t,\nu) \,,
\end{equation}
where we can see explicitly the factor of \eqref{eq:NPNS-condition}. 

This is why the $\{\CKKb,\CF,\Ccon,\Cflux\}$ terms, which do not satisfy the NPNS structure $(\lambda\neq 3)$, appear to be enhanced by $1/g_s$ in \eqref{eq:Vmin-corrected} compared to their suppression factors in the off-shell potential, while the $\{\CKKs,\Clogs,\Cxi,\Cxii\}$ terms, which do satisfy the NPNS $(\lambda = 3)$, show the same suppression in $V_{min}$ as they did in the off-shell potential \eqref{eq:Clogs}--\eqref{eq:Cxi}. 
As an explicit example, for $\{\Cflux\}$ we have
\begin{align}
    \lambda = 3 + \frac{2}{3} \,,
    && \delta V 
    \propto \Cflux\cdot\frac{g_s KM}{\V^{2/3}} \,,
    && \delta\rho 
    \propto \Cflux\cdot\frac{KM}{\V^{2/3}} \,,
\end{align}
breaking the NPNS, while for $\{\Cxii\}$ 
\begin{align}
    \lambda = 3 \,,
    && \delta V 
    \propto \Cxii \cdot g_s \,,
    && \delta\rho \propto \Cxii \cdot g_s \,.
\end{align}

\ni As a final remark, note that whether a correction breaks the NPNS structure or not, does not determine how dangerous it will be, even if it gives some information on the relative $g_s$ suppression in the off-shell potential compared to the moduli vevs and vacuum energy.  Indeed $\{\CKKs,\Clogs\}$ both have an NPNS structure, but only $\CKKs$ turns out to be safe since $\Clogs$ is anyway leading in the off-shell potential; conversely, $\{\Cflux,\CF\}$ both break the NPNS structure, but only $\CF$ turns out to be dangerous in the parameter space of weak warping. 

%% file: Sections/conclusions.tex
\section{Conclusions}
\label{sec:conclusions}

In this paper, we have studied the consequences of subleading string-loop and $\alpha'$-corrections on candidate de Sitter vacua in string effective field theories.  Consistent constructions aim to stabilise moduli, by balancing tree-level and leading-order perturbative and/or non-perturbative effects, at positive vacuum energy density, weak coupling and large volume.  Then subleading corrections, in the weak coupling and large volume expansions, may be expected to be sufficiently suppressed so as to only cause small corrections to the leading order solution.  We test this approach in the context of the recently discovered type IIB metastable de Sitter vacua that arise when an anti-D3-brane sits at the tip of a \emph{weakly-warped} conifold \cite{Bento:2021nbb}.   

The weakly-warped de Sitter vacua involve an interesting interplay between the light moduli --- the conifold deformation modulus and the volume modulus. They represent a small deviation from the famous KS/GKP/LVS solution \cite{KS2000supergravity, Giddings:2005ff,LV1}, where KS/GKP fluxes stabilise the deformation modulus and leading perturbative and non-perturbative corrections stabilise the otherwise-flat volume modulus into a scale-separated weakly-warped LVS AdS vacuum.  The deviation is sourced by the $\rm\overline{D3}$-brane, with conifold modulus $|z| \sim |z_{GKP}|\left(1- \frac{4 K \cD}{3 \pi^2 c' M \Lambda_0^4 \V^{2/3}}\right)$, and yields a total vacuum energy that is positive. The weak-warping is a previously unexplored regime of parameter space. The balance in the $\rm\overline{D3}$-brane tension with respect to the volume stabilisation, necessary to prevent volume destabilisation and usually ensured by strong-warping, is here provided by relatively large coefficients in the superpotential parameters $AW_0$.  Although this allows for solutions with low conifold flux numbers, $MK \lesssim 32$, a large $W_0$ requires a large flux contribution to the D3-tadpole \eqref{eq:DDbound}, which must therefore arise from large flux numbers in the bulk. We cannot exclude the possibility that $W_0$ can be somewhat lowered, which would result in lower $\Nflux$; however, the requirement for large tadpoles will likely remain. It is interesting that large tadpoles are also needed for control in the strongly-warped LVS \cite{Gao:2022fdi,Junghans:2022kxg}. On the other hand, the weakly-warped de Sitter solution has no bulk-singularity problem \cite{Carta_2019, Gao:2020xqh}, no problem with the conifold fitting into the bulk \cite{Carta:2021lqg} and is consistent with the Kaluza-Klein truncation, $\alpha'$ and string-loop expansions and supergravity description.

We compute the various subleading corrections that are expected to contribute to the 4d low energy effective field theory and work out their impact on the leading order metastable de Sitter vacuum.  This is summarised in Table \ref{tb:summary}. Analogous computations for the strongly-warped Large Volume Scenario have been performed recently in \cite{Junghans:2022exo} and we use the same notation as in that paper for ease of comparison.  It turns out that the parametric dependence of the subleading corrections takes the same form in the strongly-warped and weakly-warped regimes, which was not obvious a priori given that different terms dominate in the K\"ahler potential, and thus the scalar potential, in the two regimes.  The differences are then all due to the particular regime of parameter space considered.  In particular $\Cflux$ and $\Ccon$ are expected to be sensitive to the warping regime, being associated with the tip of the conifold.  Due to their parametric dependence, $\CF$ and $\CKKb$ also end up being sensitive to the warping regime.

\begin{table}[h!]
	\centering
	\def\arraystretch{1.5}
	\resizebox{\columnwidth}{!}{%
	\begin{tabular}{| c | >{\centering}m{3cm} | >{\centering\small}m{3.5cm} | c | c |} 
		\hline
		\rowcolor{black!10!white!90}  Correction & Parametric suppression & Origin & Dangerous? & c.f. strong-warping \\  
		\hline\hline
		$C^{\text{log}}_s$ & $g_s\log\mathcal{V}$ & threshold corrections to gauge couplings & yes & similar\\
		\hline
		$C^{\xi}_1$ \rule[-1em]{0pt}{3em} & $g_s\log\mathcal{V}$ & \multirow{2}{=}{\centering backreaction of 7-branes and exchange of KK modes with D7/O7 planes} & yes & similar \\
			\cline{1-2}\cline{4-5}
		$C^{\xi}_2$ \rule[-1em]{0pt}{3em} & $g_s$ & & yes & similar \\ \hline
		$C^F$ & $\frac{g_s^2 W_0^2}{\mathcal{V}^{2/3}}$ & integrating out KK modes yielding 4 superspace derivatives in EFT & yes & similar\\
				\hline\hline
		$C^{\text{KK}}_b$ & $\frac{g_s^2(g_sM)^2 \zeta^{2/3}}{\mathcal{V}^{1/3}}$ & exchange of KK modes between D7/D3 branes or O7/O3 planes & can be not & similar \\ \hline
		$C^{\text{flux}}$ & $\frac{g_s KM}{\mathcal{V}^{2/3}}$ & backreaction from fluxes, branes and O-planes that source warping & can be not & better \\ \hline \hline
		$C^{\text{KK}}_s$ & $g_s^2$ & exchange of KK modes between D7/D3 branes or O7/O3 planes & no & similar \\ \hline
		$C^{\text{con}}$ & $\frac{1}{(g_sM)^2}\frac{1}{\beta^3}$ & curvature correction in conifold region & no & better \\ \hline
		$\chi_s$ & $Ae^{-\frac{a}{g_s}\tau_s + a\frac{\chi_s}{24 g_s}}$ &  curvature corrections to D7 gauge kinetic function & no -- it helps! & better \\ \hline
	\end{tabular}
	}
	\caption{Summary of subleading corrections considered, their origin and their parametric suppression (or not) in the off-shell scalar potential, for the weakly-warped de Sitter vacuum. We also compare with the strongly-warped case, considering the danger of the corrections to candidate de Sitter vacua. 
	We do not include $\Delta\xi$ as it amounts simply to a shift in the numerical parameter $\xi$; both $\Delta\xi$ and $\chi_s$ can be computed explicitly and taken into account in the 4d EFT.  
	Note that --- in both weak and strong cases --- the LVS volume stabilisation satisfies $\log \mathcal{V} \sim 1/g_s$.}
	\label{tb:summary}
\end{table}
		
The implications for the leading order weakly-warped de Sitter vacuum --- that is, the moduli vevs  $\zeta, \mathcal{V}$ and $\tau_s$, and vacuum energy --- are as follows. The corrections $\left\{C^{\text{KK}}_s, C^{\text{con}}\right\}$ are automatically suppressed by the string-loop $g_s\ll 1$ and weak-warping $\beta\gg 1$ expansions. The corrections $\left\{C^{\text{KK}}_b, C^{\text{flux}}\right\}$ can also be parametrically suppressed consistently with the weakly-warped regime of parameter space; suppressing the $\Cflux$ correction requires a large flux contribution to the D3-tadpole, which is anyway a requirement of the parameter space with large $W_0$.  However, the $\left\{C^{\text{log}}_s,C^{\xi}_1, C^{\xi}_2\right\}$ corrections are unsuppressed due their $\log\mathcal{V}$ enhancements; this is unsurprising given that the corrections are unsuppressed in the off-shell potential and overall they would be better considered as a leading correction.  The other dangerous correction is $C^F$; it turns out that although this correction has a $g_s$ and $1/\mathcal{V}^{2/3}$ suppression, it also suffers from a $W_0$ enhancement, and overall it cannot be suppressed consistently within the weakly-warped regime that requires large $W_0 A$.  The $\chi_s$ corrections are unsuppressed, but they actually help in going to the weakly-warped regime, effectively increasing the superpotential parameter $A$. 

To summarise, we can conclude that in order for the weakly-warped de Sitter solution to be consistent, the higher dimensional construction must be such that the numerical coefficients $\left\{C^{\text{log}}_s,C^{\xi}_1, C^{\xi}_2, C^F\right\}$ are themselves suppressed or vanishing.  These are those associated with the backreaction, KK exchange and threshold corrections for D7-branes, and those associated with integrating out KK modes and the consequent derivative expansion in the 4d supersymmetric EFT. 
It would be important to understand whether or not such higher dimensional constructions are possible, both for the weakly-warped and strongly-warped constructions.\footnote{We thank Ignatios Antoniadis and Michele Cicoli for discussions on these points.} For example, the coefficient $\Clogs$ from the 1-loop K\"ahler modulus field redefinition would be loop-suppressed and small, or even absent in the case of orbifolded singularities or when $\tau_s$ does not correspond to a running gauge coupling but is instead stabilised by Euclidean D3-branes \cite{Conlon:2010ji}. The $\Cxi$ correction computed in \cite{Antoniadis:2002tr, Antoniadis:2018hqy, Antoniadis:2019rkh} arises in the presence of localised Einstein-Hilbert terms that ensure a constant Euler number for the compact Calabi-Yau in the large volume limit; such localised terms will not always be present\footnote{The $\Cxi$ correction also vanishes in the absence of local D7-tadpoles; on the other hand, the presence of local D7 tadpoles, with D7 branes not lying on O7-planes but recombining into Whitney branes, makes it easier to achieve large D3 tadpole \cite{Crino:2022zjk}.}. The numerical factors in the $\CF$ coefficients from eight-derivative corrections to the 10D action were computed in \cite{Grimm:2017okk} for examples with one K\"ahler modulus, and found to be $\mathcal{O}(10^{-4})$ (see \cite{Cicoli:2023njy} for further discussion on how $\CF$ can be small or vanishing).  As already emphasised, in addition to these small numerical coefficients, we also need large D3 tadpoles for consistency.

An interesting phenomenon that was identified in \cite{Junghans:2022exo} is the non-perturbative no-scale behaviour in some of the subleading corrections.  In detail, this occurs because in the LVS solution, the vacuum energy and one of the eigenvalues of the mass matrix turn out to have an additional $g_s$ suppression compared to the off-shell potential. Those corrections, $\{\CKKs, \Clogs, \Cxi, \Cxii, \chi_s\}$, which enter in the off-shell potential with the same LVS dependence, $e^{-\lambda\frac{a\tau_s}{g_s}}$ with $\lambda=3$, have this same $g_s$ suppression; those that break the non-perturbative no-scale behaviour, $\{\CKKb, \Cflux, \Ccon, \CF\}$  for which $\lambda \neq 3$, have one less $g_s$ suppression in the vacuum energy -- or, effectively, a $g_s$ enhancement.  We find that the non-perturbative no-scale behaviour turns out to be not decisive in whether or not corrections are dangerous; one has to recall that the NPNS $\left\{C^{\text{log}}_s,C^{\xi}_1, C^{\xi}_2\right\}$ are anyway leading in the off-shell potential with respect to the large volume expansion, and one generally has to consider the overall balance of parameters, $1/g_s$, $\mathcal{V}$, $g_s M$, $MK$ and $W_0 A$.  

Our approach in this paper has been to work out under which conditions metastable de Sitter vacua found within weakly-warped Large Volume Scenarios may be robust against expected subleading corrections.  Having established that there are well-motivated corrections that are in fact not parametrically suppressed, it would be interesting to study the corrected moduli potentials from scratch, to determine whether or not they themselves have metastable de Sitter vacua.  If they exist, they are likely not to be simply small perturbations of earlier candidate vacua and one would then have to check their own robustness against next-to-next-to-leading order corrections.  Some interesting studies on metastable de Sitter vacua in the presence of $\log\mathcal{V}$ corrections can be found for example in \cite{Antoniadis:2019rkh, Burgess:2022nbx}.  A proposal that curvature corrections on the brane could provide an alternative to warping to effectively suppress the uplifting anti-D3-brane tension was put forward in \cite{Hebecker:2022zme}.  In another direction, it would be interesting to explore the KPV instability channel \cite{Kachru:2002gs}, both leading and with corrections \`a la \cite{Hebecker:2022zme, Schreyer:2022len}, in the regime of weak-warping, where the radius of the $S^3$ at the tip is $R_{S^3} \sim \mathcal{V}^{1/6}z^{1/3}l_s$, rather than the usual $R_{S^3} \sim \sqrt{\alpha' g_sM}$.